\documentclass[aps,prb,groupedaddress,showpacs,twocolumn,floatfix]{revtex4-2}
\usepackage{mathbbol,amsmath,amsfonts,amssymb,bm,color,dsfont}
\usepackage{graphicx,psfrag,array,mathtools,lipsum,mathrsfs}
\usepackage{amsthm}
\usepackage{musicography}

\def\a{\bar a}
\def\c{\bar c}
\def\i{{\sf i}}

\begin{document}

\title{Mechanism for particle fractionalization and universal edge physics \\ in quantum Hall fluids}

\author{A. Bochniak$^{1,2}$, Z. Nussinov$^3$, A. Seidel$^3$, and G. Ortiz$^{4,}$}
\email{ortizg@iu.edu}
\affiliation{$^1$ Institute of Theoretical Physics, Jagiellonian University, \L{}ojasiewicza 11, 30-348 Krak{\'o}w, Poland \\
$^2$ Department of Mathematics, Indiana University, Bloomington, IN 47405, USA \\
$^3$  Department of Physics, Washington University in St. Louis, MO 63160, USA \\
$^4$ Department of Physics, Indiana University, Bloomington, IN 47405, USA}


\begin{abstract}
Advancing a microscopic framework that rigorously unveils the underlying topological hallmarks of fractional quantum Hall (FQH) fluids is a prerequisite for making progress in the classification of strongly-coupled topological matter. Here we advance a second-quantization framework that helps reveal an exact fusion mechanism for particle fractionalization in FQH fluids, and uncover the fundamental structure behind the condensation of non-local operators characterizing topological order in the lowest-Landau-level (LLL). We show the first exact analytic computation of the quasielectron Berry connections leading to its fractional charge and  exchange statistics, and perform Monte Carlo simulations that numerically confirm the fusion mechanism for quasiparticles. Thus, for instance, two quasiholes plus one electron of charge $e$ lead to an exact quasielectron of fractional charge $e/3$, and exchange statistics $1/3$,  in a $\nu=1/3$ Laughlin fluid. We express, in a compact manner, the sequence of (both bosonic and fermionic) Laughlin second-quantized states highlighting the lack of local condensation. Furthermore, we present a rigorous constructive subspace bosonization dictionary for the bulk fluid and establish universal long-distance behavior of edge excitations by formulating a conjecture based on the DNA, or root state, of the FQH fluid. 
\end{abstract}

\maketitle

\section{Introduction}
\label{intro}

Fractional quantum Hall (FQH)
fluids have long constituted the best known paradigm of strongly-correlated topological systems
\cite{Jain}. Nonetheless, several fundamental issues remain unresolved. These include 
the exact mechanism leading to the quasiparticle (or fractional electron) excitations and viable 
universal signatures in edge transport that are rooted in the topological characteristics of the bulk FQH fluid. 
This state of affairs is partially due to a dearth of rigorous microscopic
approaches capable of dealing with these highly entangled systems. A case in point 
is a first-principles computation of the quasielectron exchange statistics. In \cite{BCTONS}, the 
{\it Entangled Pauli Principle} (EPP) was advanced as an organizing principle for FQH 
ground states. The EPP provides information about the pattern of entanglement of the 
complete subspace of zero-energy modes, i.e., ground states, of quantum Hall parent 
Hamiltonians for both Abelian and non-Abelian fluids. Those states are generated from 
the so-called ``DNA'' \cite{BCTONS}, or root patterns \cite{Bernevig2008,ONDS}, 
which encode the elementary topological characteristics of the fluid.  

In this work we advance second-quantization many-body techniques that allow for new 
fundamental insights into the nature of quasiparticle excitations of FQH liquids. 
In particular, we present an exact fractionalization procedure that allows for a 
very natural fusion mechanism of quasiparticle generation. We determine the 
quasihole and quasiparticle operators that explicitly flesh out Laughlin's flux 
insertion/removal mechanism and provide the associated quasielectron wave function. 
The quasielectron that we find differs from Laughlin's original proposal 
\cite{laughlin}. We determine the Berry connection of this quasielectron wave function,  
considered as an Ehresmann connection on a principal fiber bundle, and as a result 
a natural fusion mechanism gets unfolded. This, in turn, leads to the exact 
determination of the quasielectron 
fractional charge. We perform Monte Carlo simulations to numerically confirm this 
fusion mechanism of fractionalization. 
In addition, we introduce an unequivocal 
diagnostic for characterizing and detecting the topological order of the FQH fluid
in terms of a {\it condensation of a non-local operator} and present 
a constructive subspace bosonization (fermionization) dictionary for
the bulk fluid that highlights the topological nature of the underlying theory. 
Our organizing EPP and the corresponding fluid's DNA encode 
universal features of the bulk FQH state and its edge excitations.  
Here we formulate a conjecture that enables a demonstration of the universal 
long-distance behavior of edge excitations in weak confining potentials. 
This is based on the exact computation of the edge Green's function over the 
DNA or root state of the topological fluid. 

Although our main results are derived in a field-theoretical manner, we will reformulate some of our 
conclusions in a first quantization language, where states become 
wave functions. For clarity, we will occasionally use a mixed representation. 

\section{States and Operator Algebra \\ in the LLL} 
\label{OperatorAlgebra}

The LLL is spanned by single-particle orbitals $\phi_r(x,y)$ 
whose functional form depends on geometry \cite{ONDS}. We consider genus zero 
manifolds such as those of the disk and the cylinder. Lengths are measured in units 
of the magnetic length $\ell=\sqrt{\frac{\hbar c}{|e| B}}$, where $B$ is the magnetic field strength, $\hbar$ the reduced Planck constant, $c$ the speed of light, and $|e|$ the magnitude
of the elementary charge. For ease of presentation, we  will primarily focus on the disk geometry \footnote{One can always apply a similarity transformation to map to the cylinder \cite{ONDS}.}. 
Then, $\phi_r(z=x+ {\sf i} y)= z^r/\mathcal{N}_r$, $\mathcal{N}_r=\sqrt{2\pi 2^r r!}$, with 
$r\ge 0$ a non-negative integer labeling the angular momentum and $z \in \mathbb{C}$ 
\footnote{Normalization is defined as $\int {\cal D}[z] (z^*)^r z^{r'} ={\cal N}_r^2 \delta_{r,r'}$ 
with ${\cal D}[z]=d^2z \, e^{-\frac{1}{2} |z|^2}$, $z^*=x-\i y$, and magnetic unit length $\ell=1$.}. $N$-particle states (elements of the Hilbert space $\mathcal{H}_{\mathrm{LLL}}$) belong to either the totally symmetric (bosons) or anti-symmetric (fermions) representations of the permutation group $S_N$. Whenever results apply to either representation, we use second-quantized creation (annihilation) $a_r^\dagger$ ($a_r^{\;}$) operators instead of the usual $c_r^\dagger$ ($c_r^{\;}$), $b_r^\dagger$ ($b_r^{\;}$), for fermions and bosons, respectively. The field operator 
\begin{eqnarray}
\Lambda^{\;}(z)=\sum_{r\ge 0} \phi_r(z) a_r^{\;}, 
\end{eqnarray}
and its adjoint $\Lambda^\dagger(z)$ satisfy canonical (anti-)commutation relations: $[\Lambda^{\;}(z),\Lambda^{\;}(z')]_\pm=0$, $[\Lambda^{\;}(z),\Lambda^{\dagger}(z')]_\pm=\{z'|z\}$,
where $\{z'|z\}=\frac{1}{2\pi}e^{\frac{zz'^\ast}{2}}$ is a bilocal kernel satisfying
$\int \mathcal{D}[z]\, \Lambda(z)\{z|z'\}=\Lambda(z')$ \cite{stone1}.
Many-particle states $|\psi\rangle \in \mathcal{H}_{\mathrm{LLL}}$ are characterized by the number 
of particles $N$ and the maximal occupied orbital $r_\mathrm{max}$, defining a filling 
factor $\nu=(N-1)/r_{\mathrm{max}}$. Given an antisymmetric 
holomorphic function $\psi$, one can construct the states 
\begin{eqnarray}
|\psi\rangle= \int \big(
\prod\limits_{i=1}^N \mathcal{D}[z_i]\big) \psi(z_1,...,z_N)\Psi^\dagger(z_1) \dots \Psi^\dagger(z_N) 
|0\rangle, \nonumber
\end{eqnarray}
in terms of the fermionic field operators $\Psi(z)$. Similarly, one can construct 
states for bosons in terms of permanents and field operators $\Phi(z)$. 

We now introduce the operator algebra necessary for the LLL 
operator fractionalization and constructive bosonization. We first review the  operator equivalents of the multivariate power-sum, $p_d(z)$, and elementary, $s_d(z)$, symmetric
polynomials ($d\ge 0$). As shown in \cite{MONS}, these are, respectively, given by
\begin{eqnarray}
\mathcal{O}_d&=&\sum_{r\ge 0}\a_{r+d}^\dagger \a^{\;}_r   \ , \ \ \mbox{and } \nonumber \\
e_d&=&\frac{1}{d!} \!\! \sum_{r_1,\dots ,r_d\ge0} \!\!\!\!
\a_{r_1+1}^\dagger  \dots \a_{r_d+1}^\dagger \a^{\;}_{r_d}\dots \a^{\;}_{r_1} 
\end{eqnarray}
(with 
$\a_r^\dagger=\mathcal{N}_r a_r^\dagger$, $\a_r^{\;}=\mathcal{N}_r^{-1} a_r^{\;}$). The operator Newton-Girard relations $de_d +\sum_{k=1}^d (-1)^k \mathcal{O}_k e_{d-k}=0$
(with $e_0=\mathds{1}$) link these operators with each other. The second-quantized extensions of the Newton-Girard relations are similar to dualities \cite{CONPRL,CON,dual^2} in that applying them twice in a row yields back the original operators. Interestingly, the operators $\mathcal{O}_d$ can be expressed in terms of Bell
polynomials 
in $e_d$'s (Appendix \ref{sec:OA}). Consequently, any quantity expressible in terms of $\mathcal{O}_d$'s can be 
also written in terms of $e_d$'s and vice versa. Both the $\mathcal{O}_d$ and $e_d$ operators generate the same commutative algebra $\mathsf{A}$. Furthermore, they satisfy the  commutation relations $[\mathcal{O}_d,\a_r]_-=-\a_{r-d}$, $[\mathcal{O}_d, 
\a_r^\dagger]_-=\a^\dagger_{r+d}$ and 
$[e_d,\a^\dagger_r]_-=\a_{r+1}^\dagger e_{d-1}$, $[e_d,\a_r]_-=-e_{d-1}\a_{r-1}$. 

A set of first-quantized symmetric operators, of relevance to Laughlin's quasielectron and conformal algebras, involves derivatives in $z$. Similar to the operators defined 
above, we introduce symmetric polynomials $p_d(\partial_z)$ and $s_d(\partial_z)$ whose second-quantized 
representations are 
\begin{eqnarray}
\mathcal{Q}_d&=&\sum_{r>d}r(r-1)\dots(r-d)\a_{r-d}^\dagger \a^{\;}_r \  , \ \mbox{ and} \nonumber \\
f_d&=&\frac{1}{d!}\sum_{r_1,\dots ,r_d>0} r_1 \dots r_d \, 
\a_{r_1-1}^\dagger  \dots \a_{r_d-1}^\dagger \a^{\;}_{r_d}\dots \a^{\;}_{r_1} , \nonumber
\end{eqnarray}
and are 
Newton-Girard-related, $d f_d +\sum_{k=1}^d (-1)^k \mathcal{Q}_k f_{d-k}=0$, 
with $f_0=\mathds{1}$. One can, analogously, define operators mixing polynomials and derivatives as in the positive ($d,d'\geq 0$) Witt algebra $[\ell_d,\ell_{d'}]_-=(d-d')\ell_{d+d'}$. These Witt algebra
generators are $\ell_d=-\sum_{i=1}^N z_{i}^{d+1}\partial_{z_i}$. Their second-quantized version is
$\hat{\ell}_d=-\sum_{r\ge 0}r \, \a_{r+d}^\dagger \a^{\;}_r$. Physically, the operators 
${\cal O}_d, e_d, \hat \ell_d$ (${\cal Q}_d, f_d$) increase (decrease) the total angular momentum or 
``add (subtract) fluxes''. Rigorous mathematical proofs appear in Appendix \ref{sec:OA}.

Symmetric operators stabilizing incompressible FQH fluids as their eigenvector with lowest eigenvalue are 
known as ``parent FQH Hamiltonians''. The EPP \cite{BCTONS,BONS} is an organizing 
principle for generating both Abelian and non-Abelian FQH states as zero modes (ground states) of frustration-free 
positive-semidefinite microscopic Hamiltonians. The Hamiltonian stabilizing Laughlin states of filling factor $\nu=1/M$, with $M$ a positive integer, is $H_{M}=\sum_m H_m$. Here  $H_m$ are the Haldane 
pseudopotentials and the sum is performed over all $0\le m<M$ sharing the (even/odd) parity of $M$.
As demonstrated in \cite{ONDS}, $H_m=\sum_{0<j<r_{\rm max}} T_{j,m}^{+}T_{j,m}^{-}$, where 
$T_{j,m}^{-}=\sum_{k} \eta_{k}(j,m) a_{j-k}a_{j+k}=(T_{j,m}^{+})^\dagger$ with 
$j=\frac{1}{2},1,\frac{3}{2},\dots,r_{\rm max}-\frac{1}{2}$ ($2k$ shares the parity of $2j$), 
and $\eta_{k}(j,m)$ are geometry-dependent form factors. For odd (even) $m$, the operator $a_r=c_r$ ($b_r$). 

The space $\mathcal{Z}_M$ of all zero modes of $H_M$ is generated by the states $|\psi\rangle\in \mathcal{H}_{\mathrm{LLL}}$ satisfying $T_{j,m}^{-}|\psi\rangle=0$. This space contains the Laughlin state $|\psi_{M}^N\rangle$ as its minimal total angular momentum, $J=MN(N-1)/2$, state. All other zero 
modes are obtained by the action of some linear combination of products of $\mathcal{O}_d$, equivalently 
$e_d$, operators onto $|\psi_{M}^N\rangle$ \cite{ONDS,MONS}. Inward squeezing is an angular-momentum preserving 
operation generated by 
\begin{eqnarray}
A^{d}_{r,r'}=\a_{r}^\dagger \a_{r'}^\dagger \a^{\;}_{r'+d}\a^{\;}_{r-d}, \ r\le r' , 
\mbox{ and } d>0 , 
\end{eqnarray}
whose multiple actions on the root partition $|\widetilde{\psi}_{M}^N\rangle=
\prod_{i=1}^N \a_{M(N-i)}^\dagger|0\rangle$ generate all occupation number eigenstates 
$|\lambda\rangle$ in the expansion of Laughlin state
$|\psi_{M}^N\rangle=|\widetilde{\psi}_{M}^N\rangle +\sum_{\lambda}C_\lambda |\lambda\rangle$, 
with integers $C_\lambda$ \cite{ONDS,MONS}. By angular momentum conservation, $\langle \psi_{M}^N|\a_r^\dagger
\a^{\;}_s|\psi_{M}^N \rangle =\alpha (r) \delta_{r,s} \|\psi_{M}^N\|^2$. In the thermodynamic limit ($N,r_{\rm max} \rightarrow \infty$ such that $\nu$
remains constant) $\alpha=N/(r_{\max}+1)\rightarrow \nu$.

\section{Operator Fractionalization and Topological Order}
\label{OperatorFractionalization} 

Our next goal is to construct second-quantized quasihole and quasiparticle operators. Following Laughlin's insertion/removal of magnetic fluxes, fractionalization is 
the notion behind that construction. Repeating this procedure $M$ times should yield an object with quantum numbers corresponding to a hole or a particle. Surprisingly, as 
we will show, the case of quasielectron excitations does not coincide with Laughlin's proposal (nor other proposals). 
As a byproduct, we will obtain a compact representation of Laughlin states (bosonic and fermionic) 
that emphasizes a sort of {\it condensation} of a non-local quantity relating to the 
topological nature of the FQH fluid. 

As shown in \cite{MONS}, the second-quantized version of the quasihole operator
$U_N(\eta)=\prod_{i=1}^N(z_i-\eta)$, $\eta\in \mathbb{C}$, is
$\widehat{U}_N(\eta)=\sum_{d=0}^N(-\eta)^{N-d}e_d$, and  satisfies 
$[\mathcal{O}_d, \widehat{U}_N(\eta)]_-=0$. Moreover \cite{MONS},
$\widehat{U}_N (\eta)\a_r^\dagger =-\eta \a_r^\dagger \widehat{U}_N (\eta) 
+\a_{r+1}^\dagger \widehat{U}_N (\eta)$ and $\a_r\widehat{U}_N(\eta)=-\eta 
\widehat{U}_N(\eta) \a_r +\widehat{U}_N(\eta) \a_{r-1}$ (see Appendix \ref{sec:OB}). The action of the quasihole operator 
on the field operator is given by \footnote{We remark that in \cite{stone1}
orbitals $\phi_r(z)$ include Gaussian factors in contrast to our convention. That implies a 
change in the differential operator to $D^{(z)}=2\partial_{z^\ast}+\frac{1}{2}z$.}
\begin{eqnarray}
\widehat{U}_N(\eta) \Lambda^\dagger(z)\!&=&\!-\eta \Lambda^\dagger(z) \widehat{U}_N(\eta)+\! \sum_{r\ge 1}
\mathcal{N}_{r-1}^{-1}\phi_{r-1}^\ast(z) \a_r^\dagger \widehat{U}_N(\eta) \nonumber \\
&=& (D^{(z)}-\eta)\Lambda^\dagger(z) \widehat{U}_N(\eta),
\end{eqnarray} 
where
$D^{(z)}=2\partial_{z^\ast}.$ The latter 
operator identity can be replaced by  $\widehat{U}_N(\eta)\Lambda^\dagger(z)\overset{\int}{=}
\left(z-\eta\right)\Lambda^\dagger(z)\widehat{U}_N(\eta)$, where the symbol $\overset{\int}{=}$ 
stresses validity following a $z$ integration \cite{stone1}.

Having established the properties of $\widehat{U}_N$, we introduce the operator
$\widehat{K}_{\mathcal{M}}(\eta)=\Lambda^\dagger(\eta) \widehat{U}_N(\eta)^\mathcal{M}$, 
for any positive integer $\mathcal{M}$. For odd  $\mathcal{M}=M$ and $\Lambda(\eta)=\Psi(\eta)$, it 
agrees with Read's non-local operator for the LLL \cite{read}. One can show that for odd (even) $\mathcal{M}$, the commutator (anticommutator) $[\widehat{K}_{\mathcal{M}}(\eta),\widehat{K}_{\mathcal{M}}(\eta')]_{-}=0$
($[\widehat{K}_{\mathcal{M}}(\eta),\widehat{K}_{\mathcal{M}}(\eta')]_{+}=0$), in the fermionic 
case, while opposite commutation relations hold for bosons. This is a consequence of the 
composite particle nature induced by the flux insertion mechanism \cite{Jain}. 

One can prove (Appendix \ref{sec:OC}) that Laughlin state can be expressed as 
\begin{eqnarray}
|\psi_{M}^N\rangle = \frac{1}{N!}{K}_{M,N-1}{K}_{M,N-2}\ldots {K}_{M,0}|0\rangle , 
\label{Laughlinstate}
\end{eqnarray}
where ${K}_{M,N}=\int \mathcal{D}[z]\, \widehat{{K}}_{M}(z)$. This indicates
that the Laughlin state does not feature a local particle condensate of ${K}_{M,N}$. This impossibility is made evident by a counting argument. Each operator ${K}_{M,N}$ adds a maximum of $M N$ units of angular momentum. Thus, a condensation of these objects would lead 
to a state with maximum total angular momentum $M N^2$. On the other hand, a state such as 
\eqref{Laughlinstate} has angular momentum $\sum_{i=0}^{N-1} M i=J$, as it should. This illustrates the above noted impossibility.

The Laughlin state, however, can be understood as a condensate of non-local objects. Consider ${\mathcal{K}}_{M}=  \int \mathcal{D}[z]\, \widehat{\mathcal{K}}_{M}(z)$ with $\widehat{\mathcal{K}}_{M}(z)=\Lambda^\dagger(z)
\widehat{\mathcal{U}}_M(z)$, and $\widehat{\mathcal{U}}_M(z)=\sum_{N\ge
0}\widehat{U}_N(z)^M|\psi_{M}^N\rangle \langle \psi_{M}^N|$ the flux-number 
non-conserving quasihole operator.  
Then, for both bosons and fermions, 
\begin{eqnarray}
|\psi_{M}^N\rangle = \frac{1}{N!}{\mathcal{K}}_{M}^N |0\rangle  . 
\end{eqnarray}
Although illuminating, this 
representation depends on $|\psi_{M}^N\rangle$ itself through $\widehat{\mathcal{U}}_M(z)$ (Appendix \ref{sec:OC}).
This condensation of non-local objects is behind the intrinsic topological order of Laughlin fluids. One 
can show this by studying the long-range order behavior of Read's operator \cite{read}. Before doing so, we need 
a result (Appendix \ref{sec:OD}) that justifies calling $\widehat{U}_N(\eta)$ the quasihole operator. Had one created 
$M$ quasiholes at position $\eta$ one should generate an object with the quantum numbers of a hole \footnote{In first-quantization language this first appeared in \cite{Girvin84} and 
an early discussion in \cite{Anderson83}. For a rigorous and general proof of Eq. (\ref{quasiholefractionalization}) see  Appendix \ref{sec:OD}.}. 
That is,
\begin{eqnarray}
\Lambda(\eta)|\psi_{M}^{N+1}\rangle = \widehat{U}_N(\eta)^M|\psi_{M}^{N}\rangle.
\label{quasiholefractionalization}
\end{eqnarray}

Studying the long-range order of Read's operator \cite{Girvin1987} amounts to establishing
that $\langle\widehat{\mathcal{K}}_M(z)^\dagger\widehat{\mathcal{K}}_M(0)\rangle$ approaches
a non-zero constant at large $|z|$ \cite{CS2019},
or alternatively,
the condensation of 
$\widehat{\mathcal{K}}_M(0)$ in the U(1) coherent state $|\theta\rangle = \sum_{N\ge
0}\sigma_{M,N} 
e^{-{\sf i} N\theta} |\psi^{N}_{M}\rangle$, where $\sigma_{M,N}=\alpha_N \|\psi_{M}^N\|^{-1}$ 
with $\alpha_N\in \mathbb{C}$ and $\theta \in \mathbb{R}$. We next expand on Read's arguments. Let us choose $\alpha_N$ such that
$\gamma_{M,N}=\sigma^\ast_{M,N}\sigma^{\;}_{M,N-1}\|\psi_M^N\|^{2}$ represents a probability
distribution concentrated around (an assumed large) $\overline{N}$. Using the
operator fractionalization relation, $\langle\theta
|\widehat{\mathcal{K}}_M(0)|\theta\rangle=e^{\i\theta}\sum_{N\ge
1}\gamma_{M,N}\|\psi_M^N\|^2\langle\psi_M^N|\Lambda^\dagger(0)\Lambda(0)|\psi_{M}^N\rangle$. 
Leading contributions to the sum come from terms with
$N$ close to $\overline{N}$, in which case $\langle
\psi_{M}^{N}|\Lambda^\dagger(0)\Lambda(0)|\psi_{M}^{N}\rangle\cong
\frac{\nu}{2\pi}\|\psi_{M}^N\|^2$. Therefore,
$\langle\theta|\widehat{\mathcal{K}}_M(0)|\theta\rangle\rightarrow \frac{\nu}{2\pi}e^{\i\theta}$
for $\overline{N}\rightarrow\infty$. Obviously, $\langle\theta|\widehat{\mathcal{K}}_M(0)|\theta\rangle$ is not a 
{\it local order parameter} \cite{read}.

Do we have a similar operator fractionalization relation for the quasiparticle operator 
$\widehat{V}_N(\eta)$, which reduces to Laughlin's quasielectron in the case of fermions?
Since within the LLL one has $\Lambda(z)\widehat{U}^\dagger_N(\eta)=(2\partial_z
-\eta^\ast)\widehat{U}^\dagger_N(\eta) \Lambda(z)$ it seems natural, by analogy to the
quasihole, to define quasiparticles as the second-quantized version of
$W_N(\eta)=\prod_{i=1}^N(2\partial_{z_i}-\eta^\ast)$, Laughlin's original proposal
\cite{laughlin}. Note, though, that the second-quantized representation of this operator 
is $\widehat{W}_N(\eta)=\sum_{d=0}^N (-\eta^\ast)^{N-d}2^d f_d$, and not
$\widehat{U}^\dagger_N(\eta)$. This proposal does not satisfy the 
operator fractionalization relation $\Lambda^\dagger(\eta) |\psi_{M}^{N-1}\rangle
=\widehat{W}_{N}(\eta)^M|\psi_{M}^{N}\rangle$ since total angular momenta do not match.
A simple modification $\Lambda^\dagger(\eta) |\psi_{M}^{N-1}\rangle
=\widehat{W}_{N-1}(\eta)^M|\psi_{M}^{N}\rangle$, can be made to match total angular momenta 
as can be easily verified by localizing the quasiparticle at $\eta=0$. A close 
inspection of the case $N=5$ shows that such a modification cannot work since,
albeit conserving the total angular momenta, individual component states display different 
angular momenta distributions (Appendix \ref{sec:OE}). A proper embodiment of the quasiparticle should satisfy \begin{eqnarray}
\Lambda^\dagger(\eta) |\psi_{M}^{N-1}\rangle \!\!&=& \!\! \widehat{V}_{N-1}(\eta)^M|\psi_{M}^{N}\rangle 
\ \ \mbox{ with} \nonumber \\
\widehat{V}_{N-1}(\eta)^M \!\! &=& \!\!\Lambda^\dagger(\eta) \widehat{U}_{N-1}(\eta)^{-M}\Lambda(\eta),
\end{eqnarray}
as can be derived from the quasihole (i.e., hole fractionalization) relation. Indeed, 
this operator is well-defined when acting on the $N$-particle Laughlin state.
Can $\widehat{V}_{N-1}(\eta)^M$ be written as the $M$-th power of another operator?
Suppose that one wants to localize a quasiparticle at $\eta=0$, then 
$\widehat{U}_{N-1}(0)^M=e_{N-1}^M$ and the problem reduces to proving that 
$\a_0^\dagger e^{-M}_{N-1}\a^{\;}_0=(\a_0^\dagger e_{N-1}^{-1} \a^{\;}_0)^{M}$. 
Recall that any Laughlin state can be obtained by an inward squeezing process 
of a root partition. Even in the bosonic case, any term in such an expansion 
has the zeroth angular momentum orbital either empty or singly occupied. 
In the first (empty) case, the action of $\a^{\;}_0$ annihilates such a term while 
in the second (singly occupied) case we are left with an $(N-1)$-particle state. The action of  
$e_{N-1}^{-1}$ on such a state reduces each remaining 
orbital component by a unit of flux. Since any such state has the smallest occupied orbital 
with $r\ge M$, the consecutive actions of $\a_0^\dagger$ and $\a^{\;}_0$ 
are well defined. It follows from the above that we can replace $\a_0^\dagger e_{N-1}^{-1} \a^{\;}_0$ by $\a_0^\dagger e_{N-1}^\dagger \a^{\;}_0$. Therefore,
\begin{equation}
\Lambda^\dagger(0) 
|\psi_{M}^{N-1}\rangle =(\Lambda^\dagger(0) \widehat{U}^\dagger_{N-1}(0)\Lambda(0))^{M}|\psi_{M}^{N}\rangle.  
\end{equation}

Analogous considerations apply to $\eta\neq 0$, as long as one can argue that the action of $\widehat{V}_{N-1}(\eta)^{k}$ is well-defined on the Laughlin state $|\psi_{M}^{N}\rangle$, for $k=1,\dotsc, M$. Indeed, if $T(\eta)$ is the magnetic translation operator by $\eta$, the translated state $T(-\eta)|\psi_{M}^{N}\rangle$ is still a zero mode of the Laughlin state parent Hamiltonian. Thus by the same squeezing argument,
$\widehat{V}_{N-1}(0)^{k}T(-\eta)|\psi_{M}^{N}\rangle$ is well-defined. Since
(up to phases) $T(\eta)\widehat{U}_{N-1}(0)T(-\eta)$ equals
$\widehat{U}_{N-1}(\eta)$, this behavior under translation carries over to $\widehat{U}_{N-1}(\eta)^{-k}$, 
$(\widehat{U}_{N-1}(\eta)^{\dagger})^k$,
and $\widehat{V}_{N-1}(\eta)^{k}$. Thus, the stated relations for the actions of these operators on the Laughlin state extend to finite $\eta$. 
We would like to stress that our quasiparticle (quasielectron) operator $\widehat{V}_{N-1}(\eta)$ does {\bf not} constitute an arbitrary Ansatz. 
It has been rigorously derived from the exact kinematic constraint that $M$ quasiparticles located at $\eta$ in an $N$-particle vacuum should be equivalent to 
the addition of one particle at the same location in an $(N-1)$-particle vacuum, i.e., the ``exact inverse" process advocated for a quasihole. From a physics standpoint, 
this constraint represents Laughlin's flux removal/insertion mechanism and is a universal property of the ground state independent of the Hamiltonian. 

\section{Quasiparticles Wave Functions} 
\label{QuasiparticleWf}

The field-theoretic approach provides an elegant formalism to prove the exact 
mechanism behind particle fractionalization. We next illustrate how this 
mechanism is translated in a first-quantized language. To this end, we start using 
a {\it mixed} representation of the quasiparticle wave function. In this 
representation the corresponding quasiparticle (quasielectron) wave function, 
localized at $\eta \in \mathbb{C}$, is given by
\begin{equation}
    \Psi^{\mathrm{qp}}_{\eta}(Z_N) = \Lambda^\dagger(\eta) \Psi_{\eta}^{(M-1)\mathrm{qh}}(Z_{N-1}),
\end{equation}
where $Z_N=\{z_1,z_2,\ldots,z_N\}$, $\Lambda^\dagger(\eta)$ 
creates a particle in the state $\psi_\eta^{0}(z)={\cal N}_0 \, 
e^{-\frac{1}{4}|z-\eta|^2}$ \footnote{In first quantization the Gaussian factor is
typically not included in the integration measure. } and
\begin{eqnarray}
    \Psi_{\eta}^{(M-1)\mathrm{qh}}(Z_{N-1})\!=\!\mathcal{N}_{\eta,N-1}^{(M-1)\mathrm{qh}}\!
    \prod\limits_{k=1}^{N-1} \! (z_k-\eta)^{M-1} \Psi_M(Z_{N-1}) \nonumber
\end{eqnarray}
is the $M-1$-quasiholes, located at $\eta$, wave function for $N-1$ particles, and
Laughlin's (un-normalized) state 
\begin{eqnarray}
    \Psi_{M}(Z_{N-1})=
    \prod\limits_{1\le i<j\le N-1}(z_i-z_j)^{M}e^{-\frac{1}{4}\sum_{k=1}^{N-1}|z_k|^2} . \nonumber
\end{eqnarray}

By the definition of the operator $\Lambda^\dagger(\eta)$, then,
\begin{equation}
    \Psi^{\mathrm{qp}}_{\eta}(Z_N) = \sqrt{N}\hat{ \mathcal{A}}\left[\psi_\eta^{0}(z_N)\Psi_{\eta}^{(M-1)\mathrm{qh}}(Z_{N-1})\right],
\end{equation}
where, for fermions for instance, 
\begin{equation}
    \hat{\mathcal{A}}\Phi(Z_N)=\frac{1}{N!}\sum\limits_{\sigma\in S_N}\mathrm{sgn}(\sigma)\Phi(z_{\sigma(1)},\ldots, z_{\sigma(N)}).
\end{equation}
This straightforwardly gives the first quantized quasiparticle wave function
\begin{equation}
\begin{split}
    \Psi^{\mathrm{qp}}_{\eta}(Z_N)=\sqrt{N}
    \mathcal{N}_{\eta,N-1}^{(M-1)\mathrm{qh}}{\cal N}_0 \, e^{-\frac{|\eta|^2}{4}}e^{-\frac{1}{4} \sum_{k=1}^N|z_k|^2}&\\
    \times \hat{\mathcal{A}}\left[e^{\frac{z_N \eta^\ast}{2}}\prod\limits_{k=1}^{N-1}(z_k-\eta)^{M-1}\prod\limits_{1\le i<j\le N-1}(z_i-z_j)^M\right]&,
    \end{split}
\end{equation}
with all normalization factors included. We claim that this wave function is 
properly normalized. Indeed, we have
\begin{equation}
    \langle \Psi^{\mathrm{qp}}_{\eta} | \Psi^{\mathrm{qp}}_{\eta}\rangle = \langle\Psi^{(M-1)\mathrm{qh}}_{\eta} |\Lambda(\eta)\Lambda^\dagger(\eta)|\Psi^{(M-1)\mathrm{qh}}_{\eta}\rangle.
\end{equation}
Since the orbital $\psi_\eta^0$ is unoccupied in $\Psi_{\eta}^{(M-1)\mathrm{qh}}$, $|\Psi_{\eta}^{(M-1)\mathrm{qh}}\rangle$ is an eigenstate of $\Lambda(\eta)\Lambda^\dagger(\eta)$ with eigenvalue $1$. Therefore,
\begin{equation}
    \langle \Psi^{\mathrm{qp}}_{\eta} | \Psi^{\mathrm{qp}}_{\eta}\rangle = \langle \Psi_{\eta}^{(M-1)\mathrm{qh}}|\Psi_{\eta}^{(M-1)\mathrm{qh}}\rangle
\end{equation}
and $\Psi_{\eta}^{\mathrm{qp}}$ is normalized if $\Psi_{\eta}^{(M-1)\mathrm{qh}}$ is normalized.

One can re-write the (un-normalized) quasiparticle (quasielectron) wave function $\bar \Psi_{\eta}^{\mathrm{qp}}$ in an enlightening manner
\begin{equation}
    \bar \Psi^{\mathrm{qp}}_\eta(Z_N)=\Gamma_\eta^\dagger(Z_N)
    \Psi_{M}(Z_N) ,
\end{equation}
with the {\it quasiparticle (quasielectron) operator}
\begin{equation}
    \Gamma_\eta^\dagger(Z_N)=\sum\limits_{i=1}^N
    e^{\frac{z_i\eta^\ast}{2}}\prod\limits_{j\neq i}\frac{(z_j-\eta)^{M-1}}{(z_j-z_i)^M},
\end{equation}
which clearly shows how it differs significantly
from prior proposals \cite{laughlin, Hansson17, Jain03, Hansson, Nielsen, Girvin86} 
(see Appendix \ref{sec:OE}).
But this is not  the whole story. It is even more illuminating to understand the 
precise mechanism leading to this remarkable quasiparticle, that we emphasize once more 
is not an Ansatz. Before doing so, 
we will first compute the charge of this excitation using the Berry connection 
idea proposed in \cite{Arovas1984} and further elaborated in Section 2.4 of 
\cite{stone1} for the quasihole, that is the Aharonov-Bohm 
effective charge coupled to magnetic flux. We will then show a remarkable exact 
property of the charge density that will shed light on the underlying 
fractionalization mechanism.

\subsection{Berry connection for one quasiparticle} 

For pedagogical reasons, we next focus on the fermionic (electron) case. 
Consider an adiabatic process (in time $t$) where the position of the quasiparticle, $\eta=\eta(t)$, is encircling an area enclosing a magnetic flux $\phi$. We will next show that the 
Berry connection decomposes into
\begin{equation}
    \left\langle \Psi^{\mathrm{qp}}_{\eta} \biggr|\frac{d}{dt} \Psi^{\mathrm{qp}}_{\eta}\right\rangle=\i \, {\cal A}_1+\i \, \tilde{\cal A}_{M-1}.
\end{equation}
As we will explain, ${\cal A}_1$ describes the Berry phase contribution from a single particle (electron) and $\tilde{\cal A}_{M-1}$ is the contribution from $M-1$ quasiholes. It is convenient to demonstrate this relation in second quantization, where only in the end, $\tilde{\cal A}_{M-1}$ is computed from first quantization methods \cite{Arovas1984, Rigolin2008,Rigolin2010}. So let $|\Psi_{\eta}^{(M-1)\mathrm{qh}}\rangle=\widehat{\psi}_\eta^\dagger |0\rangle$, where $\widehat{\psi}_\eta^\dagger$ is an element in the algebra generated by $c_j^\dagger$s, where $c_j^\dagger$ creates a particle in the orbital $\psi_0^j(z)$. Thus,
\begin{equation}
    \widehat{\psi}_\eta^\dagger=\sum\limits_{j_1,\ldots,j_{N-1}}F_{j_1,\ldots,j_{N-1}}c_{j_1}^\dagger\ldots c_{j_{N-1}}^\dagger
\end{equation}
with some coefficients $F_\bullet$ dependent on $\eta$.

The statement made earlier that $\psi_\eta^{0}(z)$ is not occupied in $|\Psi^{(M-1)\mathrm{qh}}_{\eta}\rangle$ is equivalent to saying that
\begin{equation}
\!\!\!    \Lambda(\eta)\widehat{\psi}_\eta^\dagger =(-1)^{N-1}\widehat{\psi}_\eta^\dagger\Lambda(\eta), \ \  \Lambda^\dagger(\eta)\widehat{\psi}_\eta=(-1)^{N-1}\widehat{\psi}_\eta\Lambda^\dagger(\eta).
\nonumber
\end{equation}
Trivially, also, $\Lambda(\eta)$ has the same relation with $\widehat{\psi}_\eta$, and $\Lambda^\dagger(\eta)$ (or $\frac{d}{dt}\Lambda^\dagger(\eta)$) with $\widehat{\psi}_\eta^\dagger$. From normalization,
\begin{equation}
    \Lambda(\eta)\Lambda^\dagger(\eta)|0\rangle =|0\rangle = \widehat{\psi}_\eta\widehat{\psi}_\eta^\dagger |0\rangle.
\end{equation}
Thus,
\begin{eqnarray}
\hspace*{-0.8cm}    \left\langle \Psi^{\mathrm{qp}}_{\eta} \biggr|\frac{d}{dt} \Psi^{\mathrm{qp}}_{\eta}\right\rangle &=&\left\langle 0 \biggr| \widehat{\psi}_\eta\Lambda(\eta) \left(\frac{d}{dt}\Lambda^\dagger(\eta)\right)\widehat{\psi}_\eta^\dagger \biggr|0 \right\rangle \nonumber \\
&& \hspace*{-2cm} +   \left\langle 0 \biggr| \widehat{\psi}_\eta\Lambda(\eta)\Lambda^\dagger(\eta) \left(\frac{d}{dt}\widehat{\psi}_\eta^\dagger\right) \biggr|0 \right\rangle\equiv \i \, {\cal A}_1+\i \, \tilde{\cal A}_{M-1}, 
\end{eqnarray}
where
\begin{equation}
    \begin{split}
        \i\, {\cal A}_1&=\left\langle 0 \biggr| \widehat{\psi}_\eta\Lambda(\eta) \left(\frac{d}{dt}\Lambda^\dagger(\eta)\right)\widehat{\psi}_\eta^\dagger \biggr|0 \right\rangle\\
        &=\left\langle 0 \biggr| \widehat{\psi}_\eta \widehat{\psi}_\eta^\dagger \Lambda(\eta) \left(\frac{d}{dt}\Lambda^\dagger(\eta)\right) \biggr|0 \right\rangle\\
        &=\left\langle 0 \biggr| \Lambda(\eta) \left(\frac{d}{dt}\Lambda^\dagger(\eta)\right) \biggr|0 \right\rangle=\left\langle \psi_\eta^0\biggr|\frac{d}{dt}\psi_\eta^0\right\rangle,
    \end{split}
\end{equation}
and
\begin{eqnarray}
   \hspace*{-0.1cm}     \i\, \tilde{\cal A}_{M-1} \hspace*{-0.05cm}&=&\hspace*{-0.05cm}\left\langle0\biggr| \widehat{\psi}_\eta \Lambda(\eta)\Lambda^\dagger(\eta) \left(\frac{d}{dt}\widehat{\psi}_\eta^\dagger\right)\biggr|0\right\rangle \nonumber\\
        &=&\hspace*{-0.05cm}\left\langle0\biggr| \Lambda(\eta)\Lambda^\dagger(\eta)\widehat{\psi}_\eta  \left(\frac{d}{dt}\widehat{\psi}_\eta^\dagger\right)\biggr|0\right\rangle \\
        &=&\hspace*{-0.05cm}\left\langle0\biggr|\widehat{\psi}_\eta  \left(\frac{d}{dt}\widehat{\psi}_\eta^\dagger\right)\biggr|0\right\rangle\! = \!\left\langle \Psi_{\eta}^{(M-1)\mathrm{qh}}\biggr|\frac{d}{dt}\Psi_{\eta}^{(M-1)\mathrm{qh}} \right\rangle. \nonumber
\end{eqnarray}
This finishes the proof.

Therefore, the quasiparticle charge $e^*$ has a contribution from a particle of 
charge $e$ and $M-1$ quasiholes of charge $-e/M$, i.e., $e^*=e-e (M-1)/M=e/M$, as 
expected. In simple terms, the channel fusing two quasiholes with one electron leads 
to a quasielectron of charge $e/3$ in an $\nu=1/3$ Laughlin fluid. This is a very 
intuitive (and exact) mechanism that has been overlooked until now. Notice that we 
proved that the evaluation of the quasiparticle Berry connection is exact for 
any $N$, while the quasihole charge $-e/M$ is only exact asymptotically in the 
limit $N\rightarrow \infty$ (see Section 2.4 of \cite{stone1}). 
\begin{figure*}[htb]
\centering
\includegraphics[width=1.5\columnwidth]{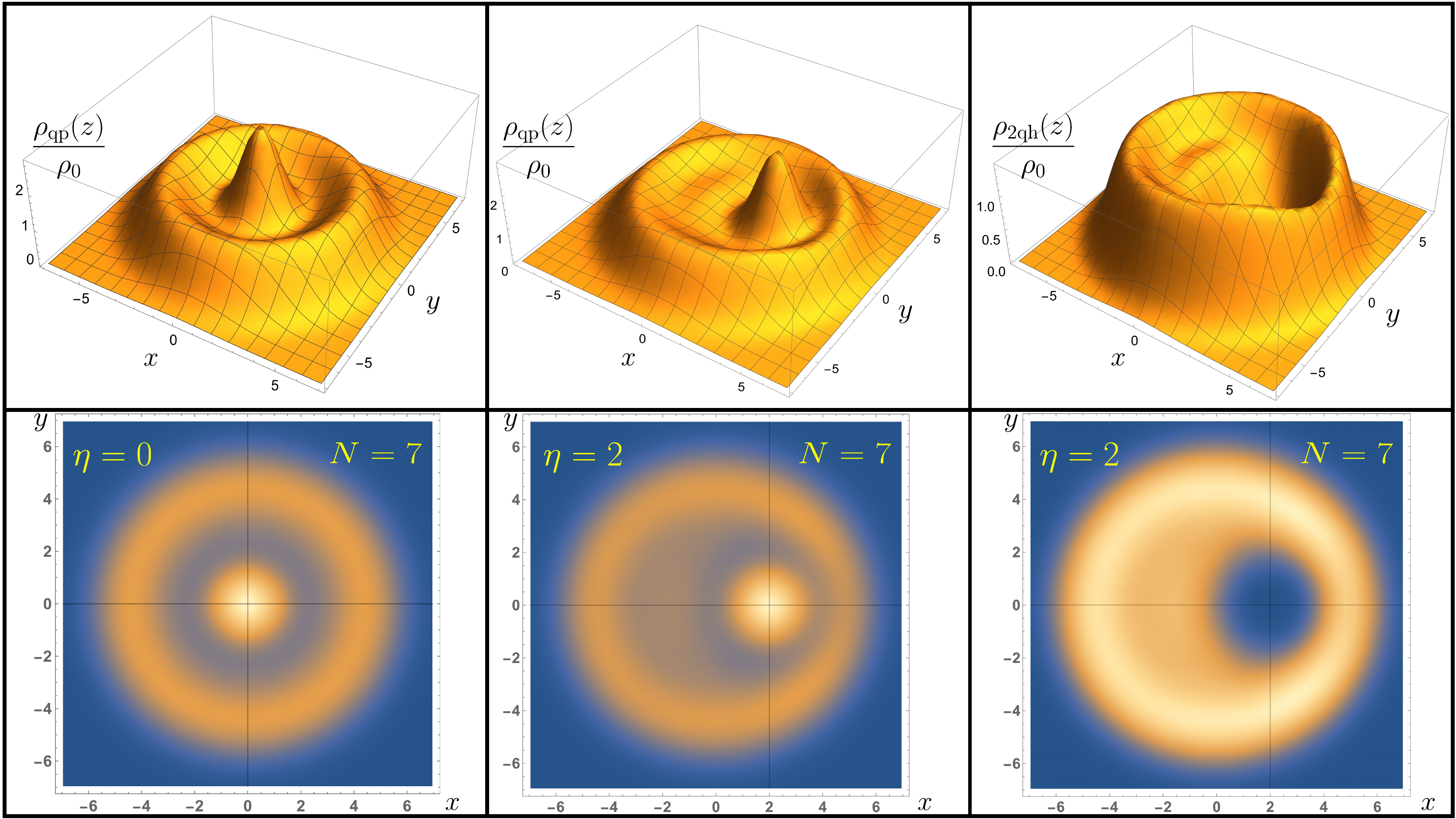}
\caption{(Color online.) Density profiles (in units of $\rho_0=\frac{\nu}{2\pi}$) 
for 1 quasielectron located at the position $\eta$ of an incompressible $\nu=\frac{1}{3}$
Laughlin fluid with $N=7$ particles (left and middle panels). The right panel depicts 
2 quasiholes in an otherwise $\nu=\frac{1}{3}$ Laughlin fluid with $N=6$ particles 
(adding the electron charge density $\frac{1}{2\pi}e^{-\frac{1}{4}|z-\eta|^2}$ leads 
to the exact same middle panel). Lower panels are contour plots of their 3D plots above. 
Monte Carlo simulations averaged over more than 2$\times 10^{10}$ equilibrated
configurations.}
\label{fig:03} 
\end{figure*}

\subsection{Charge density}

A consequence of this effective fusion mechanism manifests in the calculation of 
the quasiparticle charge density $\rho_{\mathrm{qp}}(z)$. We appeal once more to the fact that
\begin{equation}
    \Lambda(\eta)|\Psi^{(M-1)\mathrm{qh}}_{\eta}\rangle=0.
\end{equation}
This can be expressed as
\begin{equation}
\label{eq:int}
    \int d^2 r_j \psi_\eta^0(z_j)^\ast\Psi^{(M-1)\mathrm{qh}}_{\eta}(Z_{N-1})=0,
\end{equation}
for $j=1,\ldots,N-1$. Here $d^2r_j=\frac{1}{2\i}dz_j^\ast \wedge dz_j$ is the usual two dimensional 
measure on the complex plane. 

We can write the quasielectron wave function as
\begin{equation}
    \Psi^{\mathrm{qp}}_{\eta}(Z_N)=\frac{1}{\sqrt{N}}\sum\limits_{j=1}^N(-1)^j
    \psi_\eta^0(z_j) \Psi^{(M-1)\mathrm{qh}}_{\eta}(Z_{N-1},\widehat{z_j}),
\end{equation}
where $\widehat{z_j}$ means that coordinate $z_j$ is absent.

We want to evaluate
\begin{eqnarray}
    \    \rho_{\mathrm{qp}}(z) 
        &=& \! \sum\limits_{j=1}^N\int d^2 r_1 \ldots d^2 \widehat{r_j}\ldots d^2 r_N \, |\Psi^{\mathrm{qp}}_{\eta}(Z_N)|^2 \\
        =&&  \hspace*{-0.4cm} N \int d^2 r_1\ldots d^2 r_{N-1} \, |\Psi_{\eta}^{\mathrm{qp}}(Z_{N-1},z_N=z)|^2 \nonumber\\
       =&&  \hspace*{-0.4cm} \Big[
        \sum\limits_{j,j'=1}^N(-1)^{j+j'}\int d^2 r_1 \ldots d^2 r_{N-1}\psi_\eta^0(z_j)^\ast\psi_\eta^0(z_{j'})
        \nonumber \\
        \times&&  \hspace*{-0.4cm}
        \Psi_{\eta}^{(M-1)\mathrm{qh}}(Z_{N-1},\widehat{z_j})^\ast 
        \Psi^{(M-1)\mathrm{qh}}_{\eta}(Z_{N-1},\widehat{z_{j'}})
        \Big]_{z_N=z}. \nonumber
\end{eqnarray}
We now see that terms with $j\neq j'$ do not contribute. This is so since in such a case, at least one of them is not equal to $N$, say $j\neq N$, and then \eqref{eq:int} gives zero. In the $j=j'=N$ term, the integrals give a value of unity for reasons of normalization and we get
\begin{equation}
    \psi_\eta^0(z)^\ast \psi_\eta^0(z)=\langle \psi_\eta^0|\widehat{\rho}(z) 
    |\psi_\eta^0\rangle ,
\end{equation}
where $\widehat{\rho}(z)$ is the density operator. 
For $j=j'\neq N$, the integral over the $j$th variable gives $1$, and the rest precisely gives $\langle \Psi^{(M-1)\mathrm{qh}}_{\eta}|\widehat{\rho}(z) |\Psi^{(M-1)\mathrm{qh}}_{\eta}\rangle$. We have just shown that
\begin{equation}
    \rho_{\mathrm{qp}}(z)=\left\langle\psi_\eta^0|\widehat{\rho}(z) |\psi_\eta^0\right\rangle+\left\langle \Psi^{(M-1)\mathrm{qh}}_{\eta}\bigr|\widehat{\rho}(z) \bigr|\Psi^{(M-1)\mathrm{qh}}_{\eta}\right\rangle.
\end{equation}
Here, the first term is just $\frac{1}{2\pi}e^{-\frac{1}{2}|z-\eta|^2}$, while the second one is the local particle density at $z$ of $\Psi_{\eta}^{(M-1)\mathrm{qh}}(Z_{N-1})$, which is governed by a plasma analogy. 

\begin{figure}[htb]
\centering
\includegraphics[trim={0 5.2cm 0 0}, width=1.0\columnwidth]{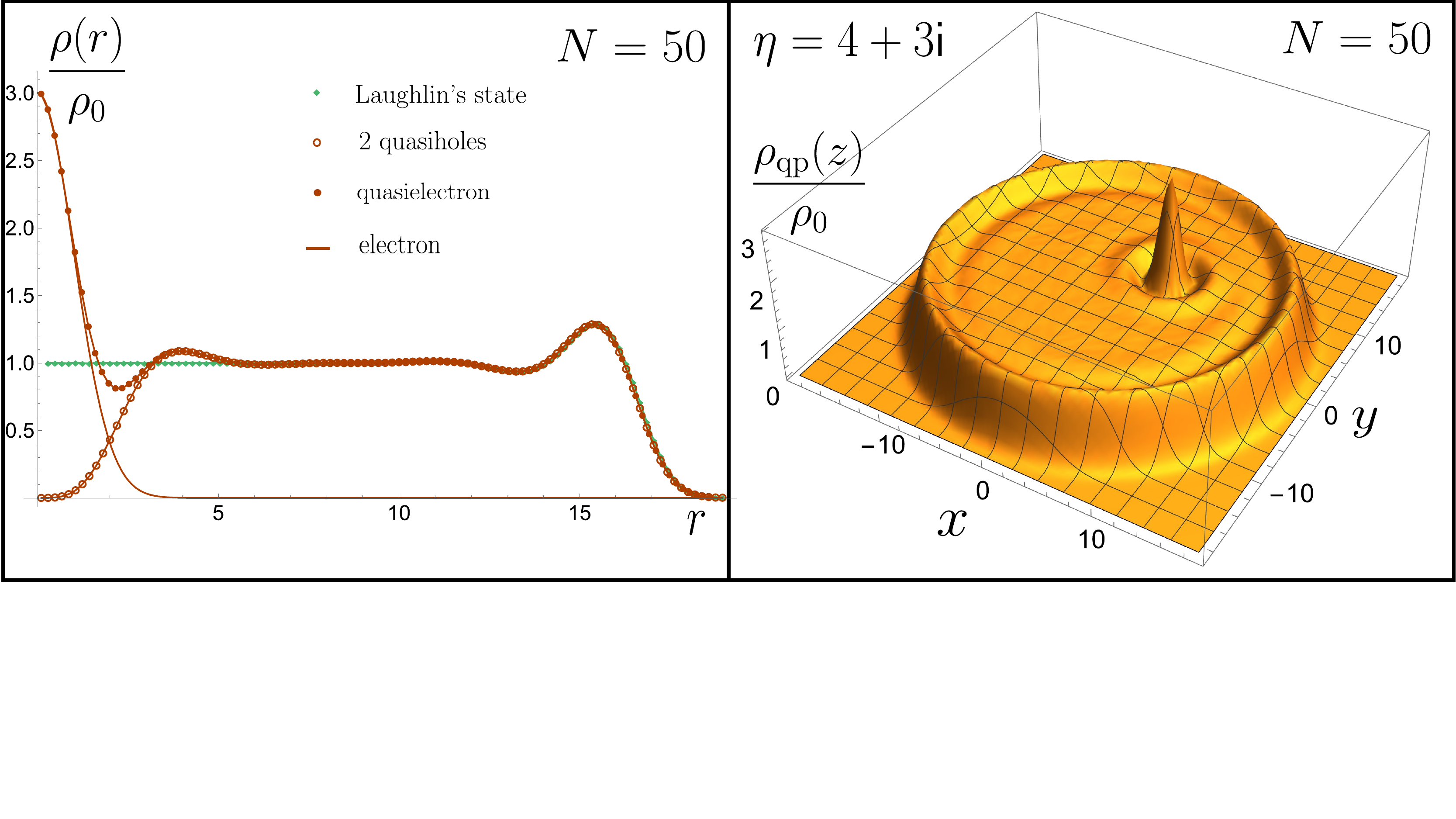}
\caption{ (Color online.) Left panel: Radial charge density $\rho(r)$ (in units of
$\rho_0=\frac{\nu}{2\pi}$) of a $\nu=1/3$ Laughlin fluid ($\rho_L(r)$ with $N=50$ 
electrons), 2 quasiholes ($N=49$ electrons), 1 electron (of density
$\frac{1}{2\pi}e^{-\frac{1}{4}|z|^2}$), and 1 quasielectron ($N=50$ electrons) 
localized at $\eta=0$. The fusion mechanism dictates that the sum of 2 quasiholes 
and 1 electron is identical to 1 quasielectron. Right panel: Quasiparticle localized at $\eta=4+3\i$. }
\label{fig:04}
\end{figure}

This picture is physically appealing. On one hand, there is no (local) plasma analogy for a 
state such as $\Psi^{\mathrm{qp}}_{\eta}(Z_N)$, but certain properties such 
as the Berry connection or the quasiparticle charge density simplify because 
of the fusion mechanism of fractionalization for any finite $N$. On the other hand, 
this same mechanism facilitate numerical computations, such as Monte Carlo \cite{Ortiz1993}, of 
certain physical properties.
For example, in Fig. \ref{fig:03} we have checked numerically that the fusion 
mechanism works for the charge density of an $N=7$ electron system and $\nu=1/3$. 
In this way, we can simulate an arbitrary large system of electrons because there 
is an "effective plasma analogy" and the Monte Carlo updates become quite efficient. 
Figure \ref{fig:04} shows Monte Carlo simulations of the radial density for a system 
of $N=50$ electrons.  
We can measure the charge of the quasiparticle by using the expression 
$\delta \rho_{\mathrm{qp}} =2\pi \int_{0}^{r_{\mathrm{cut-off}}}
\left[\rho_{\mathrm{qp}}(r)-\rho_L(r)\right]r\,dr$ where, in a finite system, 
the cut-off radius $r_{\mathrm{cut-off}}$ must at least enclose completely the  
quasiparticle and, at the same time, be sufficiently far from the boundary to avoid boundary effects \cite{KivelsonSchrieffer82}. Using the 
Monte Carlo data for $N=400$ particles (see Fig. 4 in Appendix \ref{sec:OF}) and 
choosing $r_{\mathrm{cut-off}}\leq 30 \ell$, we get a saturation of the fractional 
charge at the value $\delta\rho_{\mathrm{qp}}=0.3330(30) e$. Similarly, for two quasiholes
we get $\delta\rho_{\mathrm{2qh}}=-0.6634(30) e$.


\subsection{Berry connection for two quasiparticles: \\
The problem of statistics}

Our mechanism for particle fractionalization suggests the following form of the  wave function for a system of $N_{\mathrm{qp}}\ll N$ well-separated quasiparticles
\begin{equation}
\begin{split}
    \Psi^{N_{\mathrm{qp}}\mathrm{qp}}_{\eta_1,\ldots,\eta_{N_{\mathrm{qp}}}}(Z_N)={\cal N}^e_{\eta_1\dotsc\eta_{N_{\mathrm{qp}}}}&\Lambda^\dagger(\eta_1)\ldots \Lambda^\dagger(\eta_{N_{\mathrm{qp}}})\\
    &\times \Psi^{N_{\mathrm{qp}}(M-1)\mathrm{qp}}_{\eta_1,\ldots,\eta_{N_{\mathrm{qp}}}}(Z_{N-N_{\mathrm{qp}}}),\label{unocc}
\end{split}    
\end{equation}
where $\Psi^{N_{\mathrm{qp}}(M-1)\mathrm{qp}}_{\eta_1,\ldots,\eta_{N_{\mathrm{qp}}}}(Z_N)$ denotes the state with $M-1$ quasiholes at $\eta_1$, $M-1$ quasiholes at $\eta_2$ and so on, up to $M-1$ quasiholes at $\eta_{N_{\mathrm{qp}}}$. ${\cal N}^e_{\eta_1\dotsc\eta_{N_{\mathrm{qp}}}}$ is a normalization factor associated with the electron creation operators, as shown below.

To address the quasiparticle (a composite of one electron and $M-1$ quasiholes) exchange statistics, we next 
focus on $N_{\mathrm{qp}}=2$, in which case we get 
\begin{equation}
    \begin{split}
        &\Psi^{2\mathrm{qp}}_{\eta_1,\eta_2}(Z_N)=\sqrt{N(N-1)}\;\mathcal{N}^{2(M-1)\mathrm{qh}}_{\eta_1,\eta_2,N-2}\;
       {\cal N}^e_{\eta_1,\eta_2} \\
        &\times e^{-\frac{1}{4}\sum\limits_{k=1}^N|z_k|^2}\hat{\mathcal{A}}\left(e^{\frac{\eta_1^\ast z_N+\eta_2^\ast z_{N-1}}{2}} \prod\limits_{k=1}^{N-2}(z_k-\eta_1)^{M-1}\right. \\
        &\left.\times\prod\limits_{l=1}^{N-2}(z_l-\eta_2)^{M-1}
        \prod\limits_{1\leq i<j \leq N-2}(z_i-z_j)^M
        \right).
    \end{split}
\end{equation}

Similar to the one-particle case, $\Psi^{2(M-1)\mathrm{qh}}_{\eta_1,\eta_2}(Z_{N-2})$ has orbitals $\psi_{\eta_i}^0$, $i=1,2$, 
unoccupied, owing to the presence of factors $\prod_k (z_k-\eta_i)$, so that 
\begin{equation}
    \Lambda(\eta_i)\Lambda^\dagger(\eta_i)|\Psi^{2(M-1)\mathrm{qh}}_{\eta_1,\eta_2}\rangle =|\Psi^{2(M-1)\mathrm{qh}}_{\eta_1,\eta_2}\rangle, \quad i=1,2.
\end{equation}
By a straightforward computation, in the mixed representation, we get
\begin{equation}
\begin{split}
   &\langle  \Psi^{2\mathrm{qp}}_{\eta_1,\eta_2}| \Psi^{2\mathrm{qp}}_{\eta_1,\eta_2}\rangle =\langle \Psi^{2(M-1)\mathrm{qh}}_{\eta_1,\eta_2}|\Psi^{2(M-1)\mathrm{qh}}_{\eta_1,\eta_2}\rangle 
    ({\cal N}^e_{\eta_1,\eta_2})^2
   \\
   &\hspace{70pt}\times \left(1-\{\Lambda(\eta_1),\Lambda^\dagger(\eta_2)\}\{\Lambda(\eta_2),\Lambda^\dagger(\eta_1)\}\right)\\
  &=\langle \Psi^{2(M-1)\mathrm{qh}}_{\eta_1,\eta_2}|\Psi^{2(M-1)\mathrm{qh}}_{\eta_1,\eta_2}\rangle=1,
\end{split}  
\end{equation}
where we choose real normalization factors such that $\mathcal{N}^{2(M-1)\mathrm{qh}}_{\eta_1,\eta_2,N-2}$ normalizes the quasihole cluster state $|\Psi^{2(M-1)\mathrm{qh}}_{\eta_1,\eta_2}\rangle$ and $({\cal N}^e_{\eta_1,\eta_2})^2$
cancels the second line. The latter is just the normalization of the 2-fermion state
$\Lambda(\eta_1)^\dagger\Lambda(\eta_2)^\dagger|0\rangle$, so this choice of ${\cal N}^e_{\eta_1,\eta_2}$ can also be expressed as
\begin{equation}
({\cal N}^e_{\eta_1,\eta_2})^2
\Lambda(\eta_2)\Lambda(\eta_1)
\Lambda(\eta_1)^\dagger\Lambda(\eta_2)^\dagger|0\rangle=|0\rangle
\label{2fermnorm}
\end{equation}
and/or its Hermitian adjoint, which will be useful in the following.

For the computation of the Berry connection, just as in the one quasiparticle case, one can write $|\Psi^{2(M-1)\mathrm{qh}}_{\eta_1,\eta_2}\rangle =\hat{\psi}^\dagger_{\eta_1,\eta_2}|0\rangle$ for some $N-2$ particle operator
$\hat{\psi}^\dagger_{\eta_1,\eta_2}$ in the algebra generated by the $\Lambda(\eta)^\dagger$, in terms of which
\eqref{unocc}  can be equivalently stated as
\begin{equation}
    \Lambda(\eta_i)\hat{\psi}^\dagger_{\eta_1,\eta_2}=(-1)^{N-2}\hat{\psi}^\dagger_{\eta_1,\eta_2}\Lambda(\eta_i), \quad i=1,2.\label{unocc2}
\end{equation}
Then, utilizing the last two equation, the calculation of the Berry connection proceeds analogous to the single-particle case. In particular, one obtains two contributions 
\begin{equation}
\left\langle \Psi^{2\mathrm{qp}}_{\eta_1,\eta_2} \biggr|\frac{d}{dt} \Psi^{2\mathrm{qp}}_{\eta_1,\eta_2}\right\rangle
=\i\mathcal{A}_2+\i \tilde{\mathcal{A}}_{2(M-2)},
\end{equation}
where
\begin{equation}
  \i\mathcal{A}_2= \langle \eta_1,\eta_2|\frac{d}{dt}|\eta_1,\eta_2\rangle
\end{equation}
is the Berry connection of a normalized 2-electron state
$ |\eta_1,\eta_2\rangle=\mathcal{N}^e_{\eta_1,\eta_2}\Lambda^\dagger(\eta_1)\Lambda^\dagger(\eta_2) |0\rangle $, and
\begin{equation}
    \i\tilde{\mathcal{A}}_{2(M-2)}=\langle \Psi^{2(M-1)\mathrm{qh}}_{\eta_1,\eta_2}| \frac{d}{dt}\Psi^{2(M-1)\mathrm{qh}}_{\eta_1,\eta_2}\rangle
\end{equation}
is that of a state of two clusters of $M-1$ quasiholes.

For large $|\eta_1-\eta_2|$, both contributions are analytically under control, the 2-electron one $ \i\mathcal{A}_2$ trivially so, and the one from the quasihole cluster state,  $\i\tilde{\mathcal{A}}_{2(M-2)}$, via methods along the lines of Arovas-Schrieffer-Wilczek \cite{stone1,Arovas1984}. Dropping Aharonov-Bohm contributions, and defining the statistical phase as $e^{\i\pi\gamma}$,
the contribution to $\gamma$ from the 2-electron state is $1$ (assuming, for the time being, that the underlying particles are fermions with $M$ odd), and that of the quasihole-cluster is
$(M-1)^2/M$ \cite{Su1986}. Thus,
\begin{equation}
\pi \gamma\equiv \pi+ (M-1)^2 \cdot \frac{\pi}{M} \!\!\!\!\pmod{2 \pi} \equiv  
\frac{\pi}{M} \!\!\!\! \pmod{2 \pi},
\end{equation} as expected for a quasielectron. The same final result $\pi/M$ would be obtained for bosonic states and even $M$. 

\section{Constructive subspace bosonization} 

A bosonization map is an example of a duality \cite{CON}. Typically, dualities
are {\it dictionaries} constructed as isometries of bond algebras acting on 
the whole Hilbert space \cite{CON}. A weaker notion may involve subspaces
defined from 
a prescribed vacuum and, thus, are Hamiltonian-dependent. This is the case of
Luttinger's bosonization \cite{delft} that describes, in the thermodynamic limit, collective low energy excitations about a gapless fermion ground state. Our bosonization is performed with respect to a radically different vacuum- that of the gapped Laughlin state. Unlike most treatments, we will not bosonize the one-dimensional FQH edge (by assuming it to be a Luttinger system) but rather bosonize the entire two-dimensional FQH system.  Contrary to gapless collective excitations about the one-dimensional Fermi gas ground state associated with the Luttinger bosonization scheme, our bosonization does not describe modes of arbitrarily low finite energy but rather only the zero-energy (topological) excitations \cite{MONS} that are present in the gapped Laughlin fluid. As illustrated in \cite{ONDS,MONS}, the zero-mode subspace $\mathcal{Z}=\bigoplus_{N=0}^\infty \mathcal{Z}_N$ is generated by the action of the 
commutative algebra $\mathsf{A}$ on the Laughlin state $|\psi_{M}^N\rangle$ for different particle numbers $N$. Yet another notable difference with the conventional Luttinger bosonization (and conjectured extensions to 2+1 dimensions \cite{SSWW}) is, somewhat similar to earlier continuum renditions (as opposed to our discrete case), e.g., \cite{Fuentes}, that the indices parameterizing our bosonic excitations, $d\ge 0$, are taken from the discrete positive half-line (angular momentum values) instead of the continuous
full real line of the Luttinger system (or plane of \cite{SSWW}). Each zero-energy state in our original (fermionic/bosonic) Hilbert space has an image in the mapped
bosonized Hilbert space. Consider the following bosonic creation (annihilation) operators
$\mathfrak{b}^\dagger_d=\mathcal{O}_{d}/\sqrt{d \nu}$
($\mathfrak{b}^{\;}_d=\mathcal{O}^\dagger_{d}/\sqrt{d \nu}$). Then, $d \nu [\mathfrak{b}^{\;}_d, \mathfrak{b}^\dagger_d]_-= \sum_{r=0}^{d-1}\a_r^\dagger \a^{\;}_r$ and, in the thermodynamic limit,
\begin{equation}
 \langle \psi_{M}^N|[\mathfrak{b}^{\;}_d, \mathfrak{b}^\dagger_d]_-
|\psi_{M}^N \rangle\|\psi_{M}^N\|^{-2}\rightarrow 1.   
\end{equation}
The commutator $[\mathfrak{b}^{\;}_d, \mathfrak{b}^\dagger_{d'}]_-$ does not preserve total angular momentum 
when $d\ne d'$. It follows that, in the thermodynamic limit, within the Laughlin state subspace, $[\mathfrak{b}^{\;}_d,
\mathfrak{b}^\dagger_{d'}]_-=\delta_{d,d'}$. The field 
operator $\varphi(z)=\sum_{d\ge 0}\phi_d(z) \mathfrak{b}_d$ and its adjoint 
$\varphi^\dagger(z)$ satisfy $[\varphi(z),\varphi(z')]_-=0$ and $[\varphi(z),
\varphi^\dagger(z')]_-=\{z'|z\}$.

We next construct the operators connecting different particle sectors, that is, the 
Klein factors that commute with the bosonic degrees of freedom $\mathfrak{b}^{\;}_d, 
\mathfrak{b}^{\dagger}_d$ and are $N$-independent.
Since $|\psi_{M}^{N+1}\rangle = \frac{1}{N+1}{K}_{M,N}|\psi_{M}^N\rangle$ we define
${F}_{M,N}^\dagger=\frac{1}{N+1}{K}_{M,N}$ and $\mathcal{F}^\dagger_{M}=\sum_{N\ge 0}
{F}^\dagger_{M,N} |\psi_{M}^N \rangle \langle \psi^N_{M}|$.
This illustrates the relation between the Klein factors of bosonization with the (non-local) Read operator. We then get $\langle\psi_{M}^{N+1}|[\mathcal{O}_d,{F}^\dagger_{M,N}]_-|\psi_{M}^{N}\rangle=0$ and $\langle\psi_{M}^{N+1}|[\mathfrak{b}^{\dagger}_d,
\mathcal{F}^\dagger_{M}]_-|\psi_{M}^{N}\rangle=0$. One can prove a similar relation for 
$\mathcal{F}^{\;}_M:=(\mathcal{F}_{M}^\dagger)^\dagger$ and, analogously, for
$\mathfrak{b}^{\dagger}_d$ replaced by $\mathfrak{b}^{\;}_d$ (see Appendix \ref{sec:OG}). Since the $\widehat{U}_N(\eta)$ operators can be expressed in terms of $\mathfrak{b}_d^\dagger$'s, the fractionalization equations (both for quasiparticle as well as quasihole) can be thought of as the dictionary, at the field operator level, for our bosonization. We reiterate that this bosonization within the zero-mode subspace reflects its purely topological character. Indeed, the only Hamiltonian that commutes with the generators of $\mathsf{A}$ is the null operator. 

\section{Universal Edge Behavior} 

An understanding of the bulk-boundary correspondence in interacting 
topological matter is a long standing challenge. For FQH fluids, Wen's hypothesis \cite{Wen} for using Luttinger physics for the edge compounded by further effective edge Hamiltonian descriptions \cite{FBS,Jain01} 
constitutes our best guide for the edge physics. We now advance a conjecture enabling direct analytical calculations. We posit that the asymptotic long-distance behavior of the single-particle edge Green's function may be calculated by evaluating it for the root partition (the DNA) of the corresponding FQH state. As we next illustrate, our computed long-distance behavior shows remarkable agreement with Wen's hypothesis. Our (root pattern) angular momentum basis calculations do not include the effects of boundary confining potentials (if any exist). Most notably, we do not, at all, assume that the FQH edge is a Luttinger liquid or another effective one-dimensional system.

\begin{figure}[htb]
\centering
\includegraphics[width=\columnwidth]{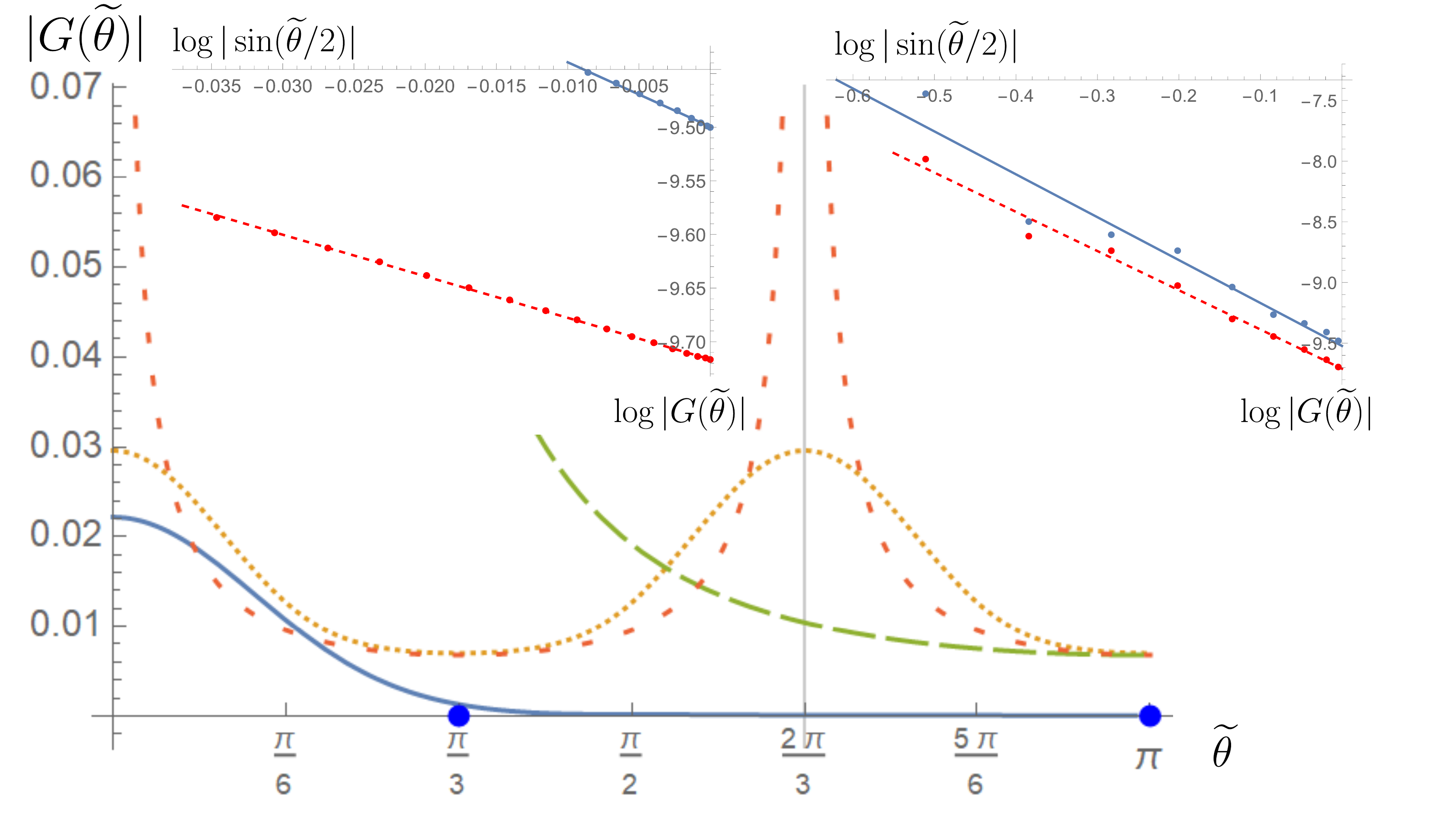}
\caption{\label{fig:01} (Color online.) Edge Green's function corresponding to the Laughlin's (blue line) and the root (orange dotted line) state, for 8 particles. Both $(\sin(\frac{\tilde{\theta}}{2}))^{-3}$ (green dashed line) and $\sin(\frac{3\tilde{\theta}}{2})$ (red dashed line) laws are also shown. The latter is a better approximation of $|\widetilde{\rho}|$ in a broader range around $\pi$. In the vicinity of blue points Stirling's approximation is valid. Insets: $\log|G(\widetilde{\theta})|$ as a function of $\log|\sin(\widetilde{\theta}/2)|$ in the range (Right): $[0.6,1]$ for $\sin(\widetilde{\theta}/2)$, for both $N=8$ (red dashed line) and $N=7$ (blue line); (Left): $[0.967,1]$ for $N=8$ (red dashed line) and $[0.991,1]$ for $N=7$ (blue line).}
\end{figure}

Consider the fermionic Green's function 
\begin{eqnarray} 
- {\sf i}G(z,z')=\rho(z,z')&=&
\langle\psi_{M}^N|\Psi^\dagger(z)\Psi(z')|\psi_{M}^N \rangle\|\psi_{M}^N\|^{-2} \nonumber
\\ 
&& \times e^{-\frac{1}{4}(|z|^2 +|z'|^2)},
\end{eqnarray} 
and coordinates $z={R}e^{{\sf i}\theta}$, 
$z'={R}e^{{\sf i}\theta'}$, where ${R}=\sqrt{2(r_{\rm max}+1)}$ is the radius of the last occupied orbital and it can be identified with the classical radius of the droplet, i.e. it satisfies $\pi R^2 \cdot \alpha=N$ with $\alpha=N/(r_{\mathrm{max}}+1)$ being the average density of the (homogeneous) droplet.
Then, 
\begin{equation}\rho(z,z')=\frac{e^{-\frac{1}{2}{R}^2}}{2\pi}
\sum_{r=0}^{r_\mathrm{max}} \left(\frac{R^2}{2}\right)^r 
\frac{e^{{\sf i}(\theta'-\theta)r}}{r!} \\ \frac{\langle \psi_{M}^N|\c_r^\dagger
\c^{\;}_r|\psi_{M}^N\rangle}{\|\psi_{M}^N\|^{2}}.\end{equation}
Similarly, the edge Green's
function associated with the root partition $|\widetilde{\psi}_{M}^N\rangle$ is 
\begin{equation}
\tilde \rho(z,z')\!= \!\frac{e^{-\frac{1}{2}{R}^2}e^{{\sf i} NM(\theta'-\theta)}}{2\pi}
\! \sum_{k=1}^{N} \! \left(\frac{{R}^2}{2}\right)^{(N-k) M}   \!\!\!\frac{ e^{{\sf i} k
M(\theta-\theta')}}{[(N-k) M]!},
\end{equation} where we used $\langle \widetilde{\psi}_{M}^N|\c_r^\dagger
\c^{\;}_r|\widetilde{\psi}_{M}^N\rangle\|\widetilde{\psi}_{M}^N\|^{-2}=1$ for
$r=0,M,\ldots, (N-1)M$, and $0$ otherwise. Thus far, our root partition calculation is exact. We next perform asymptotic analysis. For large $k$, the largest phase oscillations appear when 
$\cos(M k(\theta-\theta'))=(-1)^k$, i.e., for $|\theta-\theta'|=\tilde \theta$ near $\pi\frac{1+2l}{M}$ with $l=0,\ldots, m$ and $M=2m+1$. This implies that the dominant contributions to the sum originate from small $k$ values. We can then apply 
Stirling's approximation $\left[(N-k)M\right]!\cong\sqrt{\pi} R
\left(R^2/2\right)^{(N-k)M}e^{-R^2/2}$, 
where we 
used $R^2\nu\approx 2N$ (valid since $1-\nu \ll R^2$) leading to
\begin{equation}
\label{sM1}
     | \tilde \rho (z,z')|\cong\frac{1}{4\pi^{3/2}R \left|\sin \frac{M \tilde \theta}{2}\right|}.
\end{equation}
Long distances correspond to $\tilde \theta$ near $\pi$.
As 
\begin{eqnarray}
\left|\sin\left(\frac{M\tilde \theta}{2}\right)\right|&=&\left|\sin\left(\frac{\tilde
\theta}{2}\right)\right|^M  \\
&&-\frac{1}{8}M(M-1)(\tilde \theta-\pi)^2 + \mathcal{O}((\tilde \theta-\pi)^4), \nonumber
\end{eqnarray}
the edge Green function
\begin{equation}
\label{sM2}
    |\tilde \rho (\tilde \theta)|=\frac{1}{4\pi^{3/2}R \left|\sin\left(\frac{\tilde \theta}{2}\right)\right|^M}\left(1+\mathcal{O}((\tilde \theta-\pi)^2)\right),
\end{equation}
or, equivalently, $|\widetilde{\rho}(\widetilde{\theta})|\propto |z-z'|^{-M}$. 
This is only valid in the vicinity of $\widetilde{\theta}=\pi$ (e.g., demanding the corrections to be $\leq 1\%$, for $M=3$, restricts us to $[0.96\pi,\pi]$), while Eq. \eqref{sM1} spans a broader range -- see Fig. \ref{fig:01}. The Green's function was computed by using the tables of characters for permutation groups $S_{N(N-1)}$ for $M=3$ (up to $N=8$ and then extrapolating the results), adjusting the method in \cite{Dunne}. The difference between $|\rho|$ and $|\widetilde{\rho}|$ vanishes at $\pi$ as $N^{-1/2}$.

Nevertheless, the long-distance ($\widetilde{\theta} \sim \pi$) behavior of the Green's function, in the thermodynamic limit, cannot be reliably determined from small $N$ calculations \cite{wan}. For instance, by examining the slope $\mu$ of $\log|G(\widetilde{\theta})|$ when plotted as a function of $\log|\sin(\widetilde{\theta}/2)|$ for $N=8$ (Fig. \ref{fig:01}), we get $\mu=-3.88$ when using the range $[0.967,1]$ for $\sin(\widetilde{\theta}/2)$, while the value for $N=7$ in the range $[0.991,1]$ is $\mu=-6$. The deduced numerical value is highly dependent on the range used in the fitting procedure, e.g. for $N=8$ and the range $[0.6,1]$ we obtain $\mu=-3.23$ (for linear scale of $\widetilde{\theta}$). 
We established that the asymptotic long-distance behavior of the edge Green's function corresponding to the root state coincides with Wen's conjecture \cite{Wen}. 

\subsection{Beyond the LLL} 

The aforementioned behaviour remains true also beyond the LLL that forms the focus of our work. Indeed, repeating the above calculation when using the DNA \cite{chen17,BCTONS} of the Jain's $2/5$ state, we found $\mu=-3$, in agreement with Wen's hypothesis \cite{Wen}. In this Jain's state example, our computation captures the (EPP) entanglement structure of the root partition \cite{BCTONS} not present in Laughlin states. 

In this case we need the exact form of the following orbitals:
\begin{equation}
    \psi_{0,r}(z)=\frac{z^r}{\mathcal{N}_{0,r}}, \qquad  \psi_{1,r}(z)=\frac{{z}^*z^{r+1}-2(r+1)z^r}{\mathcal{N}_{1,r}},
\end{equation}
with $\mathcal{N}_{0,r}=\sqrt{2\pi 2^r r!}$ and $\mathcal{N}_{1,r}=\sqrt{2\pi 2^{r+2}(r+1)!}$.
The fermionic field operator is now $\Psi(z)=\sum\limits_{n,r}\psi_{n,r}(z)\overline{c}_{n,r}$ which leads to the Green's function of the following form:
\begin{equation}
\begin{split}
   \rho(z,z')=\sum\limits_{n,n'}\sum\limits_{r,r'}&\psi_{n,r}^\ast(z)\psi_{n',r'}(z')\frac{\langle \psi |\overline{c}^\dagger_{n,r}\overline{c}_{n',r'} |\psi\rangle}{\|\psi\|^2}\\
    &\times e^{-\frac{1}{4}(|z|^2+|z'|^2)},
\end{split}    
\end{equation}
where $|\psi\rangle$ is the corresponding ground state. By the angular momentum conservation, $r=r'$ under the above summation, so that
\begin{equation}
\begin{split}
   \rho(z,z')=\sum\limits_{n,n'}\sum\limits_{r}&\psi_{n,r}^\ast(z)\psi_{n',r}(z')\frac{\langle \psi |\overline{c}^\dagger_{n,r}\overline{c}_{n',r} |\psi\rangle}{\|\psi\|^2}\\
    &\times e^{-\frac{1}{4}(|z|^2+|z'|^2)},
\end{split}    
\end{equation}
For the ``DNA'' of the ground state $|\psi\rangle$ we get
\begin{equation}
\begin{split}
    \widetilde{\rho}(z,z')=\sum\limits_{n,n'}\sum\limits_{r}&\psi_{n,r}^\ast(z)\psi_{n',r}(z')\frac{\langle \mathrm{DNA} |\overline{c}^\dagger_{n,r}\overline{c}_{n',r} |\mathrm{DNA}\rangle}{\|\mathrm{DNA}\|^2}\\
    &\times e^{-\frac{1}{4}(|z|^2+|z'|^2)}.
\end{split}
\end{equation}
The ``DNA'' of the ground state of the Jain's $2/5$ state \cite{BCTONS,chen17} is of the form $|\mathrm{DNA}\rangle = \prod\limits_{k\ge 0} \widehat{\varphi_{5k+3}}|0\rangle$ with
%
\begin{equation}
\begin{split}
    \widehat{\varphi_r}=\alpha_{0,0}(r)&\left(\overline{c}_{0,r-1}^\dagger \overline{c}_{0,r+1}^\dagger +\frac{\sqrt{r+2}}{2}\overline{c}_{0,r-1}^\dagger \overline{c}_{1,r+1}^\dagger\right.\\
    &\left.-\frac{\sqrt{r}}{2}\overline{c}_{1,r-1}^\dagger \overline{c}_{0,r+1}^\dagger\right),
\end{split}    
\end{equation}
where $\alpha_{0,0}$ is an $r$-dependent factor.

As a result, 
\begin{equation}
\label{rholl}
    \begin{split}
        &\widetilde{\rho}(z,z')=\frac{e^{-\frac{1}{4}(|z|^2+|z'|^2)}}{2\pi}\\
        &\times\left\{
        \sum\limits_{0\le 5l+2\le r_{\mathrm{max}}}\frac{({z}^*z')^{5l+2}}{2^{5l+2}h_l(5l+2)!}\left[1+(5l+5)^2\right.\right.\\
        &\left.\left.+ \frac{2(5l+3)-|z'|^2}{4} + (2(5l+3)-|z|^2)(5 l +3)\right]\right.\\
        &\left.+\sum\limits_{0\le 5l+4\le r_{\mathrm{max}}}\frac{({z}^*z')^{5l+4}}{2^{5l+4}h_l(5l+4)!}\left[1+(5l+3)^2\right.\right.\\
        &\left.\left.+ \frac{|z'|^2-2(5l+5)}{4} + (|z|^2-2(5l+5))(5 l +5)\right]
        \right\},
    \end{split}
\end{equation}
where $h_l=1+(5l+3)^2+(5l+5)^2$.

Henceforth, we will focus on points $z=R e^{\i \theta}$ and $z'=R e^{\i \theta'}$ that lie on the edge. 
We next discuss the two contributions to $\widetilde{\rho}(z,z')$ in Eq. (\ref{rholl}).

We start by discussing the contribution to $\widetilde{\rho}(z,z')$ of exponent $5l+2$. 
With the above polar substitution for the boundary points $z$ and $z'$, this  becomes
\begin{equation}
\begin{split}
    &\frac{e^{-\frac{1}{2}R^2}}{2\pi}\sum\limits_{k=1}^{\mathcal{N}}\left(\frac{R^2}{2}\right)^{5(\mathcal{N}-k)+2}\frac{
    e^{-\i(5(\mathcal{N}-k)+2)(\theta-\theta')}
    }{h_{\mathcal{N}-k}\left(5(\mathcal{N}-k)+2\right)!}\\
    &\times \left[1+ \left(5(\mathcal{N}-k)+5\right)^2\right.
    \\
    &\left.+\frac{2\left(5(\mathcal{N}-k)+3\right)-R^2}{4}\cdot (20(\mathcal{N}-k)+13)\right] ,
\end{split}    
\end{equation}
where $\mathcal{N}=\lfloor \frac{1}{5}\left(\frac{N-1}{\nu}-2\right)\rfloor+1$ with $\nu=\frac{2}{5}$, and we have used the same change of summation index as in the case of the LLL.

Assume now that only small integer $k$ are of relevance in the above summation - we will check validity of this assumption later on. 
Then using the Stirling approximation, the fact that $N\cong \frac{R^2 \nu}{2} \gg 1$ and $\mathcal{N}-k\cong \mathcal{N}$, we get
\begin{equation}
    \left[5(\mathcal{N}-k)\right]!\cong \sqrt{\pi} R \left(\frac{R^2}{2}\right)^{5(\mathcal{N}-k)+2} e^{-\frac{R^2}{2}},
\end{equation}
and, as a result, for the part with the exponent $5l+2$, we end up with
\begin{equation}
    \frac{e^{-\i5\mathcal{N}(\theta -\theta')}}{2R \pi^{3/2}}\kappa_2(\mathcal{N},R)\sum\limits_{k=1}^{\mathcal{N}} e^{\i(5k-2)(\theta -\theta')},
\end{equation}
where $\kappa_2(\mathcal{N},R)$ is a certain rational function in $\mathcal{N}$.

Similar to the above, for the part having $5l+4$ as an exponent, we get
\begin{equation}
    \frac{e^{-\i5\widehat{\mathcal{N}}(\theta-\theta')}}{2R \pi^{3/2}}\kappa_4(\widehat{\mathcal{N}},R)
    \sum\limits_{k=1}^{\widehat{\mathcal{N}}} e^{\i(5k-4)(\theta-\theta')}
\end{equation}
with a rational, in $\widehat{\mathcal{N}}=\lfloor\frac{1}{5}\left(\frac{N-1}{\nu}-4\right)\rfloor+1$, function $\kappa_4(\widehat{\mathcal{N}},R)$.

Next, we observe that for large radius $R$ we can without the loss of generality assume that $\widehat{\mathcal{N}}=\mathcal{N}$, so that $e^{-\i 5\widehat{\mathcal{N}}(\theta-\theta')}$'s lead to an irrelevant global factor since at the very end we will be interested in the absolute value of the Green's function. We now argue that in the thermodynamic limit, $\frac{\kappa_2}{\kappa_4}\rightarrow 1$. Indeed, since $R^2\sim \frac{2N}{\nu}$ and $\nu=\frac{2}{5}$ we have $R^2\sim 5N$. Moreover, we know that $\mathcal{N}\sim \frac{N}{5\nu}\sim \frac{N}{2}$. Hence $R^2\sim 10\mathcal{N}$. This shows that $\frac{\kappa_2}{\kappa_4}\rightarrow 1$. Furthermore, this also shows that, in the limit $\mathcal{N}\rightarrow \infty$, we have for $\widetilde{\rho}$:
\begin{equation}
    \frac{e^{-\i 5\mathcal{N}(\theta -\theta')}}{4R \pi^{3/2}}\left( \sum\limits_{k=1}^{\infty} e^{\i(5k-2)(\theta-\theta')}+\sum\limits_{k=1}^\infty e^{\i(5k-4)(\theta-\theta')}\right),
\end{equation}
since both $\kappa_2$ and $\kappa_4$ tends to $\frac{1}{2}$ in this limit. 

We next explain why the assumption $\mathcal{N}-k\cong \mathcal{N}$ is valid. Towards this end, one needs to verify that the only $k$ values that matter are the small ones, i.e., that $\cos((5k-2)(\theta-\theta'))\cong(-1)^k$. This is indeed true (in particular around $\theta-\theta'=\pi$, which is exactly our point of interest). Analogous considerations work also for the term of exponent $5l+4$. Therefore, the problem reduces to the evaluation of
\begin{equation}
\begin{split}
    &\frac{1}{4R \pi^{3/2}}\left|\sum\limits_{k=1}^\infty e^{\i(5k-2)(\theta-\theta')}+e^{\i(5k-4)(\theta-\theta')}\right|\\
    &=\frac{1}{4R \pi^{3/2}}\left| \frac{\i e^{\i \frac{3}{2}(\theta-\theta')} \cos(\theta-\theta')}{\sin\left(\frac{5 (\theta-\theta')}{2}\right)}   \right|.
    \label{eq:0}
\end{split}    
\end{equation}
To ascertain the long distance behavior, we examine angular differences $|\theta-\theta'|=\tilde \theta$ close to $\pi$, where this asymptotically becomes
\begin{equation}
\begin{split}
    \frac{1}{4R \pi^{3/2}}\left|\frac{1-2\sin^2\left(\frac{\widetilde{\theta}}{2}\right)}{\sin\left(\frac{5\widetilde{\theta}}{2}\right)}\right|\cong \frac{1}{2R \pi^{3/2}}\frac{1}{\left|\sin\left(\frac{\widetilde{\theta}}{2}\right)\right|^3}.
\end{split}    
\end{equation}
The above derived result is in agreement with Wen's conjecture \cite{Wen} for the FQH Jain's $2/5$ liquid.



\section{Conclusions} 

Our approach sheds light on the elusive exact mechanism underlying fractionalized 
quasielectron excitations in FQH fluids (and formalizes the fractionalization of quasihole
excitations \cite{Girvin84}). By solving an outstanding open problem \cite{Hansson,Nielsen}, 
our construct underscores the importance of a systematic operator based microscopic approach
complementing Laughlin's original quasiparticle wave function Ansatz. The algebraic structure 
of the LLL is deeply tied to the Newton-Girard relations. We have shown that there are 
numerous pairs of ``dual'' operators that are linked to each other via these relations
(including operators associated with the Witt algebra). The Newton-Girard relations typically convert a local operator into a non-local ``dual'' operator. A main message of the present work 
is that ``derivative operations'' on FQH vacua do not lead to exact quasiparticle excitations. The precise mechanism leading to charge fractionalization consists 
of the joint process of flux and (original) particle insertions. In other words, an elementary 
fusion channel of quasiholes and an electron generates a quasielectron excitation. For 
instance, to generate one quasielectron excitation in a $\nu=1/M$  Laughlin fluid one 
needs to insert $M-1$ fluxes, in an ($N-1$)-electron fluid, and fuse them with an additional
electron. A fundamental difference between quasihole and quasiparticle excitations 
can be traced back to their $M$-clustering properties \cite{BCTONS}. While quasiholes 
preserve the $M$-clustering property of the incompressible (ground state) fluid, 
quasiparticle states breaks it down. This is at the origin of the asymmetry between 
these two kinds of excitations. Equivalently, while quasihole wave functions sustain 
a (local) plasma analogy this is not the case for quasielectrons. 
 
We explicitly {\it constructed the quasiparticle (quasielectron) wave function}. Our found fusion mechanism of quasiparticle generation is not only 
the mathematically exact (for an arbitrary number of particles) field-theoretic operator 
procedure but it is also behind the 
exact analytic computation of other quasiparticle properties, such as its charge density and Berry 
connections leading to the right fractional charge and exchange statistics. This is a truly unprecedented remarkable result that 
we have numerically confirmed via detailed Monte Carlo simulations.

Intriguingly, within our field-theoretic framework, we find that the Laughlin state is a condensate of a non-local Read type operator. Our approach allows for a constructive (zero-energy) subspace bosonization of the full two-dimensional system that further evinces the non-local topological character of the problem and, once again, cements links to Read's operator. The constructed Klein operator associated with this angular momentum (and flux counting) root state based bosonization scheme is none other than Read's non-local operator. We suspect that this angular momentum (flux counting) based mapping might relate to real-space flux attachment (and attendant Chern-Simons) type bosonization schemes \cite{fradkin,SSWW}. Lastly, we illustrated how the long-distance behavior of edge excitations associated with the root partition component (DNA) of the {\it bulk} FQH ground state may be readily calculated. Strikingly, the asymptotic long-distance edge physics derived in this manner matches Wen's earlier hypothesis in the cases that we tested. This agreement hints at a possibly general powerful {\it computational recipe for predicting edge physics}.

\section{Acknowledgements}   

We thank J. Jain, H. Hansson, and S. Simon for useful comments. G.O. acknowledges
support from the US Department of Energy grant DE-SC0020343. A.B. acknowledges the Polish-US 
Fulbright Commission for the possibility of visiting Indiana University, Bloomington, during 
the Fulbright Junior Research Award scholarship. Work by A.S. has been supported by the 
National Science Foundation Grant No. DMR-2029401.

\bibliography{pnas.bib}

\appendix

\section{Newton-Girard dual pairs}
\label{sec:OA}

Let $d\ge 0$, and consider operators $\mathcal{O}_d$ and $e_d$, defined for both fermions and bosons. Since they are operator analogues of the symmetric polynomials $p_d(z)=\sum\limits_{i=1}^N z_i^d$ and $s_d(z)=\sum\limits_{1\le i_1\le \ldots \le i_d\le N}z_{i_1}\ldots z_{i_d}$, respectively, one may expect that the operator version of the Netwon-Girard relation, known from the theory of symmetric polynomials, holds also in this case:
\begin{equation}
    de_d +\sum\limits_{k=1}^d (-1)^k \mathcal{O}_k e_{d-k}=0.
    \label{eq:NG}
\end{equation}
Indeed, this was already proven in \cite{MONS}. Then, $\mathcal{O}_d$ operators can be expressed recursively in terms of $e_d$'s: 
\begin{equation}
    \mathcal{O}_d=(-1)^{d-1}\left(de_d + \sum\limits_{k=1}^{d-1}(-1)^k \mathcal{O}_k e_{d-k}\right),
\end{equation}
so that any operator $\mathcal{O}_d$ can be represented in terms of $e_d$'s only, and vice versa. Moreover, this can be achieved by a (many-variable) polynomial relation. In order to write an explicit expression for $\mathcal{O}_d$ in terms of $e_d$'s operators we start with the Newton-Girard relation, \eqref{eq:NG}, and for any positive $d$ we get
\begin{equation}
    \begin{cases}
     e_1=\mathcal{O}_1,\\
     2e_2=e_1\mathcal{O}_1-\mathcal{O}_2,\\
     \vdots\\
     de_d=e_{d-1}\mathcal{O}_1-e_{d-2}\mathcal{O}_2+\ldots + (-1)^de_1\mathcal{O}_{d-1} \\ \hspace*{4.56cm } + (-1)^{d-1}\mathcal{O}_d.
    \end{cases}
\end{equation}

Since, again by the Newton-Girard relation, every coefficient which is a multiple of $e_k$ is also in the algebra $\mathsf{A}$ of $\mathcal{O}_d$ operators, this leads to a system of $d$ linear equations in the commutative algebra $\mathsf{A}$. We would like to solve this linear system for unknowns $(\mathcal{O}_1,-\mathcal{O}_2,\mathcal{O}_3,\ldots,(-1)^d\mathcal{O}_{d-1},\mathcal{O}_d)$. By Cramer's theorem applied to the ring $\mathsf{A}$, the necessary and sufficient condition for the existence of the unique solution is the invertibility in $\mathsf{A}$ of the determinant of the matrix encoding this system. In our case this is a lower triangular matrix with diagonal $(\underbrace{1,\ldots, 1}_{d-1}, (-1)^{d-1})$, so its determinant, equal to $(-1)^{d-1}$, is clearly an invertible element in $\mathsf{A}$. Therefore, again by  Cramer's theorem, we get (see also \cite[p.~28]{macdonald}):
\begin{eqnarray}
    \mathcal{O}_d&=&(-1)^{d-1}\det \!\!
    \begin{pmatrix} 1 & 0 & 0&\ldots & 0&e_1\\
    e_1 & 1 & 0&\ldots & 0& 2e_2 \\
    e_2 & e_1 & 1 &\ldots & 0&3e_3\\
    \vdots & \vdots & \vdots & \ddots & \vdots & \vdots\\
    e_{d-2}& e_{d-3}& e_{d-4}&\ldots & 1& (d-1)e_{d-1}\\
    e_{d-1} & e_{d-2} & e_{d-3} & \ldots & e_1 & de_d
    \end{pmatrix} \nonumber \\
    &=&\det\begin{pmatrix}
    e_1 & 1& 0 & 0 &\ldots &0\\
    2e_2 & e_1 &1 &0 &\ldots &0\\
    3e_3& e_2 & e_1 &1 &\ldots &0\\
    \vdots&\vdots&\vdots&\vdots & \ddots & \vdots\\
    de_e& e_{d-1} &e_{d-2}& e_{d-3} & \ldots &e_1
    \end{pmatrix} . \nonumber
\end{eqnarray}
Expanding the above determinant one can finally find
\begin{eqnarray}
    \mathcal{O}_d&=&(-1)^d d \sum\limits_{\substack{r_1+2r_2+\ldots + dr_d=d\\
    r_1,\ldots,r_d\ge 0}}\frac{(r_1+\ldots + r_{d}-1)!}{r_1!\ldots r_2!} \nonumber \\
    && \hspace{3cm} \times (-1)^{r_1+\ldots + r_d}e_1^{r_1}\ldots e_d^{r_d} \nonumber \\
    &=&(-1)^d d\sum\limits_{k=1}^d\frac{1}{k}\hat{B}_{d,k}(-e_1,\ldots,-e_{d-k-1}),
\end{eqnarray}
where $\hat{B}_{n,k}(x_1,\ldots,x_{n-k+1})=\frac{k!}{n!}B_{n,k}(1!x_1,2!x_2,\ldots,(n-k+1)!x_{n-k+1})$ and $B_{n,k}$ is the (exponential) Bell polynomial \cite{comtet}.

Let now $f_d$ be the operator representing, in the second quantization picture, the elementary symmetric polynomial with the $z_i$ variables substituted by partial derivatives, $s_d(\partial_z)$, and let $\mathcal{Q}_d$ be the second-quantized version of the operator $p_d(\partial_z)$. Since the Newton-Girard relation holds for $s_d(\partial_z)$ and $p_d(\partial_z)$, an analogous one is expected for the pair of operators $f_d$ and $\mathcal{Q}_d$. 


First, we claim that the second-quantized version $f_d$ of the differential operator $s_d(\partial_z)$ is 
\begin{equation}
    f_d=\frac{1}{d!}\sum\limits_{r_1,\ldots, r_d> 0}r_1\ldots r_d\, \overline{a}_{r_{1}-1}^\dagger \ldots \overline{a}_{r_d -1}^\dagger \overline{a}_{r_d}\ldots \overline{a}_{r_1}.
\end{equation}

We start with the fermionic case. For $\beta_1>\ldots >\beta_N\geq 0$ consider the monic monomial $r_\beta(Z_N)=z_1^{\beta_1}\ldots z_N^{\beta_N}$, where  $\beta=(\beta_1,\ldots,\beta_N)$ and $Z_N=(z_1,\ldots,z_N)$. $r_\beta$ is a homogeneous polynomial of degree $|\beta|=\beta_1+\ldots +\beta_N$. Let 
$R_\beta(Z_N)=N!\widehat{\mathcal{A}} r_\beta(Z_N)$ be defined by the total antisymmetrization $\widehat{\mathcal{A}}$ of $r_\beta$ in variables forming $Z_N$. This polynomial corresponds, in second quantization, to the operator $\overline{\mathfrak{c}}_\beta^\dagger=\overline{c}_{\beta_1}^\dagger\ldots \overline{c}_{\beta_N}^\dagger$, the Slater determinant. 

Notice that for $d>N$, $f_d\overline{\mathfrak{c}}_\beta=0$, and similarly $s_d(\partial_z)R_d(Z_N)=0$. Therefore, below we implicitly assume that $d$ is at most $N$. We notice that
\begin{eqnarray}
\label{eq:fd}
        s_d(\partial_z)R_\beta(Z_N)&=&N!\widehat{\mathcal{A}}s_d(\partial_z)r_\beta(Z_N) \\
&&\hspace*{-2.5cm}        = N!\widehat{\mathcal{A}}\mathcal{S}_{(i_1,\ldots i_d)}\left(\partial_{z_{i_1}}\ldots \partial_{z_{i_d}}z_1^{\beta_{1}}\ldots z_N^{\beta_N}\right)  \nonumber \\
&&\hspace*{-2.5cm}        =N!\widehat{\mathcal{A}}\mathcal{S}_{(i_1,\ldots i_d)}\left(\beta_{i_1}\ldots \beta_{i_d}z_1^{\beta_1}\ldots z_{i_1}^{\beta_{i_1}-1}\ldots z_{i_{d}}^{\beta_{i_d}-1}\ldots z_N^{\beta_N}\right)  \nonumber \\
&&\hspace*{-2.5cm}        =\mathcal{S}_{(i_1,\ldots i_d)}\left(\beta_{i_1}\ldots \beta_{i_d} R_{(\beta_1,\ldots, \beta_{i_1}-1,\ldots, \beta_{i_d-1},\ldots, \beta_{N})}(z)\right). \nonumber
\end{eqnarray}
Here $\mathcal{S}_{(i_1,\ldots, i_d)}$ denotes the total symmetrization in $i_1,\ldots, i_d$ with $i_j\neq i_k$ for every pair $(j,k)$ of indices $j,k\in\{1,\ldots,d\}$ such that $j\neq k$.

Observe that when acting by the operator $f_d$ on the state $\overline{\mathfrak{c}}_\beta^\dagger|0\rangle$ the symmetrization $\mathcal{S}_{(i_1,\ldots i_d)}$ is explicitly involved. Moreover, each term under the symmetrization produces 
\begin{equation}
\label{state:perm}
    \beta_{i_1}\ldots \beta_{i_d}\overline{c}_{\beta_{i_1}-1}^\dagger \ldots \overline{c}_{\beta_{i_d}-1}^\dagger \overline{c}_{\beta_{i_d}}\ldots \overline{c}_{\beta_{i_1}}\overline{c}_{\beta_1}^\dagger\ldots \overline{c}_{\beta_{N}}^\dagger|0\rangle.
\end{equation}
To finish the proof it is enough to show that when placing the fermionic operators in the canonical order no sign is generated out of the reordering permutation. By counting signs appearing in the reordering, one can easily check that this is indeed the case. 

For bosons, in the definition of $R_\beta(Z_N)$ we replace the total antisymmetrization by a symmetrization, which corresponds to having a permament instead of a Slater determinant. The reasoning in \eqref{eq:fd} follows with only minor adjustments. Now $\mathcal{S}_{(i_1,\ldots, i_d)}$ denotes the total symmetrization in $i_1,\ldots, i_d$ without the additional assumption that the indices are pairwise distinct, contrary to the fermionic case. Obviously, no sign counting in the reordering is needed in the bosonic case. 

In order to justify that $\mathcal{Q}_d=\sum\limits_{r>0}r(r-1)\ldots (r-d)\overline{a}_{r-d}^\dagger\overline{a}_r$ is the second-quantized version of the $p_d(\partial_z)$ polynomial, we will show that this operator satisfies the Newton-Girard equation. First notice that mimicking the computation from \cite{MONS} one can easily prove that
\begin{equation}
\label{comm_f}
    [f_d,\overline{a}_r^\dagger]_-=r\overline{a}_{r-1}^\dagger f_{d-1}, \qquad [f_d,\overline{a}_r]_-=-(r+1)f_{d-1}\overline{a}_{r-1},
\end{equation}
for any $d>0$ and $r\ge 0$. Moreover, we have
\begin{equation}
    \begin{split}
        f_d&=\frac{1}{d}\sum\limits_{r\ge 0} r\overline{a}^\dagger_{r-1} f_{d-1}\overline{a}_r.
    \end{split}
\end{equation}
Next, iterating this equation sufficiently many times we find ($f_1=\mathcal{Q}_1$)
\begin{equation}
    \begin{split}
        df_d&=\sum\limits_{r\ge0} r\left(f_{d-1}\overline{a}_{r-1}^\dagger - (r-1)\overline{a}_{r-2}^\dagger f_{d-2}\right)\overline{a}_r\\
        &=f_{d-1}\mathcal{Q}_1 - \sum\limits_{r\ge 0}r(r-1)\overline{a}_{r-2}^\dagger f_{d-2}\overline{a}_r\\
        &=f_{d-1}\mathcal{Q}_1 \!- \! \sum\limits_{r\ge0} r(r-1) \!\! \left(f_{d-2}\overline{a}_{r-2}^\dagger -(r-2)\overline{a}_{r-3}^\dagger f_{d-3}\right)\\
        &=f_{d-1}\mathcal{Q}_1 +(-1)f_{d-2}\mathcal{Q}_2 \\ 
        & \hspace*{1.6cm} +(-1)^2 \sum\limits_{r\ge 0}r(r-1)(r-2) \overline{a}_{r-3}^\dagger f_{d-3}\\
        &=\ldots = -\sum\limits_{k=1}^d (-1)^k f_{d-k}\mathcal{Q}_k.
    \end{split}
\end{equation}
By induction, each $f_d$ commutes with every $\mathcal{Q}_k$, so that we end up with the Newton-Girard relation for the operators $f_d$ and $\mathcal{Q}_k$:
\begin{equation}
    df_d +\sum\limits_{k=1}^d (-1)^k \mathcal{Q}_k f_{d-k}=0.
\end{equation}


\subsection{The Witt algebra}

For $d\geq 0$ we consider also the differential operators $\ell_d=-\sum\limits_{i=1}^N z_{i}^{d+1}\partial_{z_i}$. They satisfy $[\ell_d,\ell_{d'}]_-=(d-d')\ell_{d+d'}$, i.e. form the positive Witt algebra $\mathfrak{W}_+$. Another set of operators satisfying the same algebra is given by $\widehat{\ell}_d=-\sum\limits_{r\ge 0}r \overline{a}_{r+d}^\dagger \overline{a}_r$. We claim that $\widehat{\ell}_{d}$ is the second quantization version of $\ell_d$. Indeed, we will show it explicitly for fermions and left the adjustments in the bosonic case to the reader as they are exactly the same as we discussed above in the case of the $f_d$ operator. We have $\ell_d r_\beta(Z_N)=-\sum\limits_{i=1}^N \beta_i z_1^{\beta_1}\ldots z_i^{\beta_i+d}\ldots z_N^{\beta_N}$. Since $\ell_d$ is symmetric under transpositions $(i,j)$ for any two indices $i,j$ under the summation, from the very definition of the totally antisymmetrized polynomial $R_\beta(Z_N)$ we have also that $\ell_d R_\beta(Z_N)=N! \widehat{\mathcal{A}}\ell_d r_\beta(Z_N)$, and, as a result, we end up with $\ell_d R_\beta(Z_N)=-\sum\limits_{j=1}^N \beta_j R_{(\beta_1,\ldots, \beta_{j-1},\beta_{j}+d,\beta_{j+1},\ldots, \beta_N)}(Z_N)$. 
On the other hand, for the Slater determinant corresponding to the polynomial $R_\beta(Z_N)$ we have $\widehat{\ell}_d \overline{\mathfrak{c}}_\beta^\dagger |0\rangle=\sum\limits_{j=1}^N \beta_j \overline{c}_{\beta_1}^\dagger \ldots \overline{c}_{\beta_{j-1}}^\dagger \overline{c}_{\beta_{j}+d}^\dagger \overline{c}_{\beta_{j+1}}^\dagger \ldots \overline{c}_{\beta_N}^\dagger |0\rangle$, where we have used the fact that the total number of transpositions required to properly order the creation operators is even. It remains to notice that the Slater determinant $c_{\beta_1}^\dagger \ldots \overline{c}_{\beta_{j-1}}^\dagger \overline{c}_{\beta_{j}+d}^\dagger \overline{c}_{\beta_{j+1}}^\dagger \ldots \overline{c}_{\beta_N}^\dagger$ corresponds to the polynomial $R_{(\beta_1,\ldots, \beta_{j-1},\beta_{j}+d,\beta_{j+1},\ldots, \beta_N)}(Z_N)$. 

\section{Properties of the quasihole operator}
\label{sec:OB}

The quasihole operator $\widehat{U}_N(\eta)=\sum\limits_{d=0}^N(-\eta)^{N-d}e_d$ satisfies the following important relations:
\begin{equation}
    \widehat{U}_N(\eta)\overline{a}_r^\dagger = -\eta\overline{a}_r^\dagger \widehat{U}_N(\eta)+\overline{a}_{r+1}^\dagger \widehat{U}_N(\eta),
\end{equation}
\begin{equation}
    \overline{a}_r\widehat{U}_N(\eta)=-\eta \widehat{U}_N(\eta)\overline{a}_r+\widehat{U}_N(\eta) \overline{a}_{r-1}.
\end{equation}

Since the first of those relations was already demonstrated in \cite{MONS}, we present here, for completeness, 
the proof of the second one. This follows as a result of a straightforward computation:
\begin{equation}
    \begin{split}
        \widehat{U}_N(\eta)\overline{a}_r&=(-\eta)^{\widehat{N}} \sum\limits_{d\ge 0} (-\eta)^{-d}e_d \overline{a}_r\\
        &=(-\eta)^{\widehat{N}}\sum\limits_{d\ge 0}(-e_{d-1}\overline{a}_{r-1}+\overline{a}_{r}e_d)\\
        &=-(-\eta)^{\widehat{N}}(-\eta)^{-1}\sum\limits_{d\ge 0}(-\eta)^{-d-1}e_{d-1}\overline{a}_{r-1} \\
        &\hspace*{1cm} +\overline{a}_r(-\eta)^{-1}(\eta)^{\widehat{N}}\sum\limits_{d\ge0}(-\eta)^{-d}e_d\\
        &=\frac{1}{\eta}\widehat{U}_N(\eta)\overline{a}_{r-1}-\frac{1}{\eta}\overline{a}_r\widehat{U}_N(\eta).
    \end{split}
\end{equation}

\section{Laughlin sequence states}
\label{sec:OC}

We start with the following

{\bf Lemma~1.}
The $M$th power of the quasihole operator is given by 
\begin{equation}
    \widehat{U}_N(z)^M=(-1)^{MN}\sum\limits_{r\ge 0}z^r \widehat{S}^{\musSharp{}}_{MN-r},
\end{equation}
where 
$\widehat{S}^{\musSharp{}}_{l}=(-1)^l\sum\limits_{n_1+\dots+n_M=l}e_{n_1}\dots e_{n_M}$ for $l>0$, and $0$ 
otherwise, as defined in \cite{CS15,CS2019}.

Before we proceed with the proof we remark that our $M$ corresponds to $M+1$ in \cite{CS15,CS2019}, and  
we are working with the disk geometry as opposed to the infinitely thick cylinder 
geometry used therein. Moreover, 
we use the phase $(-1)^{MN}$ which is appropriate both for fermions and bosons.

\begin{proof} First, we notice that
\begin{equation}
    \begin{split}
        \widehat{U}_N(z)^M&=\left(\sum\limits_{d=0}^N (-z)^{N-d}e_d\right)^M\\
        &=\sum\limits_{k_0+\ldots + k_N=M} (-z)^{Nk_0 + (N-1) k_1 +\ldots+ 1\cdot k_1 + 0\cdot k_0} \\
        &\hspace*{1.5cm} \times\binom{M}{k_0,\ldots,k_N} e_0^{k_0}\ldots e_N^{k_N}.
    \end{split}
\end{equation}
Next, we observe that
\begin{equation}
    \begin{split}
    & N k_0+ (N-1) k_1 +\ldots +1\cdot k_{N-1} + 0\cdot k_N \\
    &= N(k_0+\ldots + k_N) - (0\cdot k_0 +1\cdot k_1 +\ldots + N\cdot k_N )
    \\
    &= MN-(0\cdot k_0 +1\cdot k_1 +\ldots + N\cdot k_N ),
    \end{split}
\end{equation}
where we use the constraint $k_0+\ldots + k_N=M$ imposed under the summation. Therefore, this leads to
\begin{eqnarray}
\hspace*{-1.0cm}
    \widehat{U}_N(z)^M=\sum\limits_{k_0+\ldots+k_N=M}(-z)^{NM-(0\cdot k_0 + \ldots + N\cdot k_N)} \nonumber \\
       \hspace*{1.4cm} \times\binom{M}{k_0,\ldots,k_N} e_0^{k_0}\ldots e_N^{k_N}
\end{eqnarray}
Continuing along these lines we get
\begin{equation}
    \begin{split}
        \widehat{U}_N(z)^M&=\sum\limits_{r\ge 0}\sum\limits_{k_0+\ldots+k_N=M}\delta_{r,MN - (0\cdot k_0 + \ldots + N\cdot k_N)} (-z)^r \\
        &\hspace*{1.5cm} \times\ \binom{M}{k_0,\ldots,k_N} e_0^{k_0}\ldots e_N^{k_N}\\
        &=\sum\limits_{r\ge 0} z^r (-1)^r\sum\limits_{k_0+\ldots+k_N=M}\delta_{MN-r,0\cdot k_0 + \ldots + N\cdot k_N}\\
        &\hspace*{1.5cm} \times\ \binom{M}{k_0,\ldots,k_N} e_0^{k_0}\ldots e_N^{k_N}\\
        &=\sum\limits_{r\ge 0} z^r (-1)^r \sum\limits_{n_1+\ldots +n_M=MN-r}e_{n_1}\ldots e_{n_M}\\
        &=(-1)^{MN}\sum\limits_{r\ge 0} z^r \widehat{S}^{\musSharp{}}_{MN-r}.
    \end{split}
\end{equation} 
\end{proof}

From the above Lemma,
\begin{equation}
 {K}_{M,N}=(-1)^{MN} \sum_{r\ge 0} \a_r^\dagger 
\widehat{S}^{\musSharp{}}_{MN-r},   
\end{equation}
after integration over $z$, and together with the recurrence relation \cite{CS15,CS2019},
\begin{equation}
 |\psi_{M}^N\rangle=\frac{(-1)^{M(N-1)}}{N} \sum_{r\ge 0}\overline{a}_r^\dagger \widehat{S}^{\musSharp{}}_{M(N-1)-r}|\psi_{M}^{N-1}\rangle,  
\end{equation} 
results in 
\begin{equation}
 |\psi_{M}^N\rangle =\frac{1}{N} {K}_{M,N-1}|\psi_{M}^{N-1}\rangle.   
\end{equation} 


We now prove by induction that $|\psi_M^N\rangle=\frac{1}{N!}\mathcal{K}_M^N|0\rangle$. Indeed, we have
\begin{equation}
\begin{split}
    \frac{1}{N!}\mathcal{K}_M^N|0\rangle &=\left(\frac{1}{N}\mathcal{K}_M\right)\left(\frac{1}{(N-1)!}\mathcal{K}_{M}^{N-1}|0\rangle\right) \\
    &\hspace*{-.9cm}  =\frac{1}{N}\mathcal{K}_M|\psi_{M}^{N-1}\rangle\\
    &\hspace*{-.9cm}  = \sum\limits_{N'\ge 0} \frac{1}{N}\int \mathcal{D}[z]\Lambda^\dagger(z)\widehat{U}_{N'}(z)^{M}|\psi_M^{N'}\rangle\langle\psi_M^{N'}|\psi_M^{N-1}\rangle\\
    &\hspace*{-.9cm}  =\frac{1}{N}K_{M,N-1}|\psi_{M}^{N-1}\rangle=|\psi_M^N\rangle. 
\end{split}
\end{equation}
\section{Quasihole operator fractionalization}
\label{sec:OD}

One of the consequence of the Lemma~1 is the operator relation satisfied by the quasihole,
\begin{equation}
    \Lambda(\eta)|\psi_{M}^{N+1}\rangle = \widehat{U}_N(\eta)^M|\psi_{M}^{N}\rangle.
\end{equation}
This can be proven with the help of Lemma~1 and the 
following identity
\begin{equation}
    \Lambda(z)|\psi^{N+1}_{M}\rangle = (-1)^{MN} \sum_{r\ge 0}z^r 
\widehat{S}^{\musSharp{}}_{MN-r} |\psi^{N}_{M}\rangle.
\end{equation}

\section{The quasiparticle (quasielectron) problem}
\label{sec:OE}

Here we elaborate on the quasiparticle definition within the framework of second quantization. First, we stress that the operator $f_d$ differs from $e_d^\dagger$ exactly by the factors $r_1\ldots r_d$ in each term under the summation. This is one of the reasons behind the differences between the various quasiparticle (quasielectron) wave function proposals. Let us concentrate for convenience on the quasielectron case. Interestingly, one can check by an explicit computation that both these operations produce exactly the same (up to a global factor) polynomials out of the ones representing Laughlin $3-$ and $4-$particle states. Moreover, they also agree (again, up to a global factor) with the action of $(\overline{c}_0^\dagger e_2^{-1}\overline{c}_0)^3$ and $(\overline{c}_0^\dagger e_3^{-1}\overline{c}_0)^3$, respectively \footnote{The analogous statement is in those cases also true for higher values of $M$.}. The first difference can be observed for $N=5$, where in $\Psi^\dagger(0)|\psi_3^4\rangle$ the highest occupied orbital is equal to nine, while the action by $f_4^3$ on $|\psi_3^5\rangle$ produces terms with the twelfth orbital occupied. This issue is caused by the lack of presence of the $\overline{c}_0$ and $\overline{c}_0^\dagger$ operators. Moreover, the numerical coefficients also differ as a result of the presence of additional $r_1\ldots r_d$ factors in the second-quantized version of the $s_d(\partial_z)$ operator.


\subsection{Comparison with other proposals}

Our approach is field-theoretic and, therefore, allows an algebraic proof of operator 
fractionalization. There are, however, other first-quantization approaches proposing 
corrections to Laughlin's quasielectron state. It is fair, then, to compare all these 
proposals. To this end, in the following we'll consider the case of a quasielectron localized 
at the center of a disk ($\eta=0$) in a $\nu=\frac{1}{3}$ QH fluid.

\subsubsection{Laughlin's quasielectron}

The (holomorphic part of the) wave function for Laughlin's quasielectron is
\begin{equation}
   \Psi_0^L(Z_N)=\prod_{i=1}^N(2\partial_{z_i})\prod^N\limits_{\substack{j<k}}(z_j-z_k)^3.
\end{equation}
In particular, for $N=3$ this leads to the second-quantized form
\begin{equation}
    |\Psi_0^L(Z_3)\rangle=(-288\overline{c}_5^\dagger\overline{c}_1^\dagger\overline{c}_0^\dagger+720\overline{c}_4^\dagger\overline{c}_2^\dagger\overline{c}_0^\dagger-2880\overline{c}_3^\dagger \overline{c}_2^\dagger \overline{c}_1^\dagger)|0\rangle.
\end{equation}
The angular momenta counting leads to the conclusion that Laughlin's quasielectron operator cannot describe fractionalization. Indeed, we can observe this effect directly here by applying the operator $\prod\limits_{i=1}^3 (2\partial_{z_i})$ three times on the $3$-particle Laughlin state. In this case we get immediately zero, which is obviously not the required result, $\Psi^\dagger(0)|\psi_3^2\rangle$. A modification  of Laughlin's original proposal could, in principle (due to the angular momenta match), work, but as already mentioned, the difference can be easily found already in case of $N=5$.

\subsubsection{Approach based on conformal block}

The (holomorphic part) of the quasielectron wave function resulting from the CFT approach \cite{Hansson17} is given by
\begin{equation}
   \Psi_0^{\mathrm{CFT}}(Z_N)=\sum\limits_{i=1}^N(-1)^i\prod^N\limits_{\substack{j<k \\ j,k\neq i}}(z_j-z_k)^3\, \partial_{z_i}\prod^N\limits_{l\neq i}(z_l-z_i)^2,
\end{equation}
since a single CFT quasielectron localized at zero agrees with Jain's approach \cite{Jain03} based on composite fermions \cite[p.~37 -- discussion below Eqn. (70)]{Hansson17}.

In this case, again for $N=3$, we get the second-quantized form
\begin{equation}
    |\Psi_0^{\mathrm{CFT}}(Z_3)\rangle=(5\overline{c}_5^\dagger\overline{c}_1^\dagger \overline{c}_0^\dagger -10\overline{c}_4^\dagger \overline{c}_2^\dagger \overline{c}_0^\dagger +40\overline{c}_3^\dagger\overline{c}_2^\dagger\overline{c}_1^\dagger)|0\rangle.
\end{equation}
One can also try to compare with the wave function proposed in \cite[Eq.~(11)]{Hansson} for a quasielectron in a state with angular momentum one, in order to match the angular momenta counting. In such a case we end up with
\begin{equation}
    |\Psi_0^{\mathrm{CFT};\,(1)}(Z_3)\rangle = (2\overline{c}_5^\dagger\overline{c}_2^\dagger \overline{c}_0^\dagger -10\overline{c}_4^\dagger\overline{c}_3^\dagger \overline{c}_0^\dagger +10\overline{c}_4^\dagger \overline{c}_2^\dagger \overline{c}_1^\dagger)|0\rangle.
\end{equation}

\subsubsection{Our quasielectron} 

In first quantization (the holomorphic part of) our quasielectron wave function is 
\begin{equation}
   \Psi_0^{\rm qe}(Z_N)=(-1)^N\sum\limits_{i=1}^N(-1)^i\prod^N\limits_{\substack{j<k \\ j,k\neq i}}(z_j-z_k)^3 \prod^N\limits_{l\neq i}z_l^{2} .
   \label{eqqel}
\end{equation}
Indeed, from our second-quantized prescription for the quasielectron we deduce that 
\begin{equation}
    \Psi_0^{\rm qe}(Z_N)=\sum\limits_{j=1}^N\frac{\psi_3^N(z_1,\ldots,z_{j-1},0,z_{j+1},\ldots, z_N)}{z_1\ldots\widehat{z_j}\ldots z_N},
\end{equation}
where $\widehat{z_j}$ means that the variable $z_j$ is not present in the product. This expression immediately leads to \eqref{eqqel}.

The action of our quasielectron operator, $\bar c_0^\dagger e_2^\dagger \bar c_0$, on the Laughlin's $3$-particle state $|\psi_3^3\rangle$ produces the second-quantized state
\begin{equation}
  |\Psi_0^{\rm qe}(Z_3)\rangle=(\overline{c}_5^\dagger \overline{c}_2^\dagger\overline{c}_0^\dagger-3\overline{c}_4^\dagger \overline{c}_3^\dagger \overline{c}_0^\dagger)|0\rangle,
\end{equation}
which differs from all the proposals above. Moreover, we notice that applying the operator $\overline{c}_0^\dagger e_2^\dagger \overline{c}_0$ three times on the state $|\psi_3^3\rangle$ give us $-3\overline{c}_2^\dagger \overline{c}_1^\dagger \overline{c}_0^\dagger|0\rangle$, which is exactly the action of $\overline{c}_0^\dagger =\Psi^\dagger(0)$ on Laughlin state $|\psi_3^2\rangle$, as required by  fractionalization. 

Moreover, at general quasielectron location $\eta$, we can present the quasielectron wavefunction in the following mixed first/second quantization form:

\begin{equation}
  \Psi_\eta^{\rm qe}(Z_N)=
\Lambda(\eta)^\dagger\,
\cfrac{1}{\prod\limits_{i=1}^{N-1} (z_i-\eta)} \,\Lambda(\eta) 
\,\psi_M^N(Z_N) .
\end{equation}
In particular, the squeezing arguments of the main text imply that despite the rational factors, the resulting wave function is still analytic up to Gaussian factors, i.e., is in the LLL.

\subsubsection{MacDonald-Girvin's quasielectron}

Let us start with the observation that the action of the operator $e_N^\dagger$ on an $N$-particle state $|\psi\rangle$ that does not have the zeroth orbital occupied is equivalent to the action of $\mathsf{D}=\prod_{r\ge 1}\hat d_r$ with $\hat d_r=1-\c_r^\dagger \c^{\;}_r+\c_{r-1}^\dagger \c^{\;}_r$, and the operators $\hat{d}_r$ are ordered according to their index $r$, with $\hat{d}_1$ acting first. Since $|\psi\rangle$ has always well-defined maximal occupied orbital $r_{\mathrm{max}}$, this product is always finite, with the upper limit $r_{\mathrm{max}}$, since $\hat d_{r>r_\mathrm{max}}$ acts as an identity operator on $|\psi\rangle$. In order to prove that these two actions coincide take any term $\c_{\beta_1}^\dagger \ldots \c_{\beta_N}^\dagger |0\rangle$ in the decomposition (into Slater determinants) of $|\psi\rangle$. Notice that by our assumption $\beta_j>0$ for any $j=1,\ldots, N$. Then, $N! e^\dagger_{N}\c_{\beta_1}^\dagger \ldots \c_{\beta_N}^\dagger |0\rangle=\c_{\beta_1-1}^\dagger \ldots \c_{\beta_N-1}^\dagger |0\rangle$. On the other hand, consider consecutive actions of $\hat d_1,\ldots, \hat d_{r_{\mathrm{max}}}$ on $|\psi\rangle$. By induction, $\hat d_r$ acting on $\c_{\beta_1}^\dagger \ldots \c_{\beta_N}^\dagger |0\rangle$ leaves it unchanged if for all $j=1,\ldots, N,\ \beta_j\neq r$, or reduces $\beta_j$ by $1$ otherwise (since, by induction, $\beta_j-1$ was unoccupied before this action). Therefore, consecutive actions of $\hat d_1,\ldots, \hat d_{r_{\mathrm{max}}}$ on $|\psi\rangle$ is equivalent to $e_{N}^\dagger |\psi\rangle$ (up to an irrelevant $N!$ factor).

Our quasielectron state localized at $\eta=0$ is $\c_0^\dagger  e_{N-1}^\dagger \c^{\;}_0 |\psi_M^N\rangle$, and notice that $\c^{\;}_0 |\psi_M^N\rangle$ is an $N-1$-particle state which does not have occupied the zeroth orbital. Therefore, the discussion from the previous paragraph applies and we have 
\begin{equation}
    \c_0^\dagger  e_{N-1}^\dagger \c^{\;}_0 |\psi_M^N\rangle=\c_0^\dagger  \mathsf{D} \c^{\;}_0 |\psi_M^N\rangle.
\end{equation}
Since $\c_0^\dagger$ commutes with $\hat d_r, \ r\ge 1$, we can write the above state in the form $\mathsf{D} \c_0^\dagger \c^{\;}_0 |\psi_M^N\rangle$.

In \cite{Girvin86} a quasielectron state was proposed to be of the form $\mathsf{D}(1-\c_0^\dagger \c^{\;}_0) |\psi_M^N\rangle$, which differs from the one we derived
in the current work. These authors have only provided an Ansatz for a quasiparticle located at $\eta=0$.

\section{Numerical computation of fractional charge}
\label{sec:OF}

\begin{figure}[b]
\centering
\includegraphics[width=0.9\columnwidth]{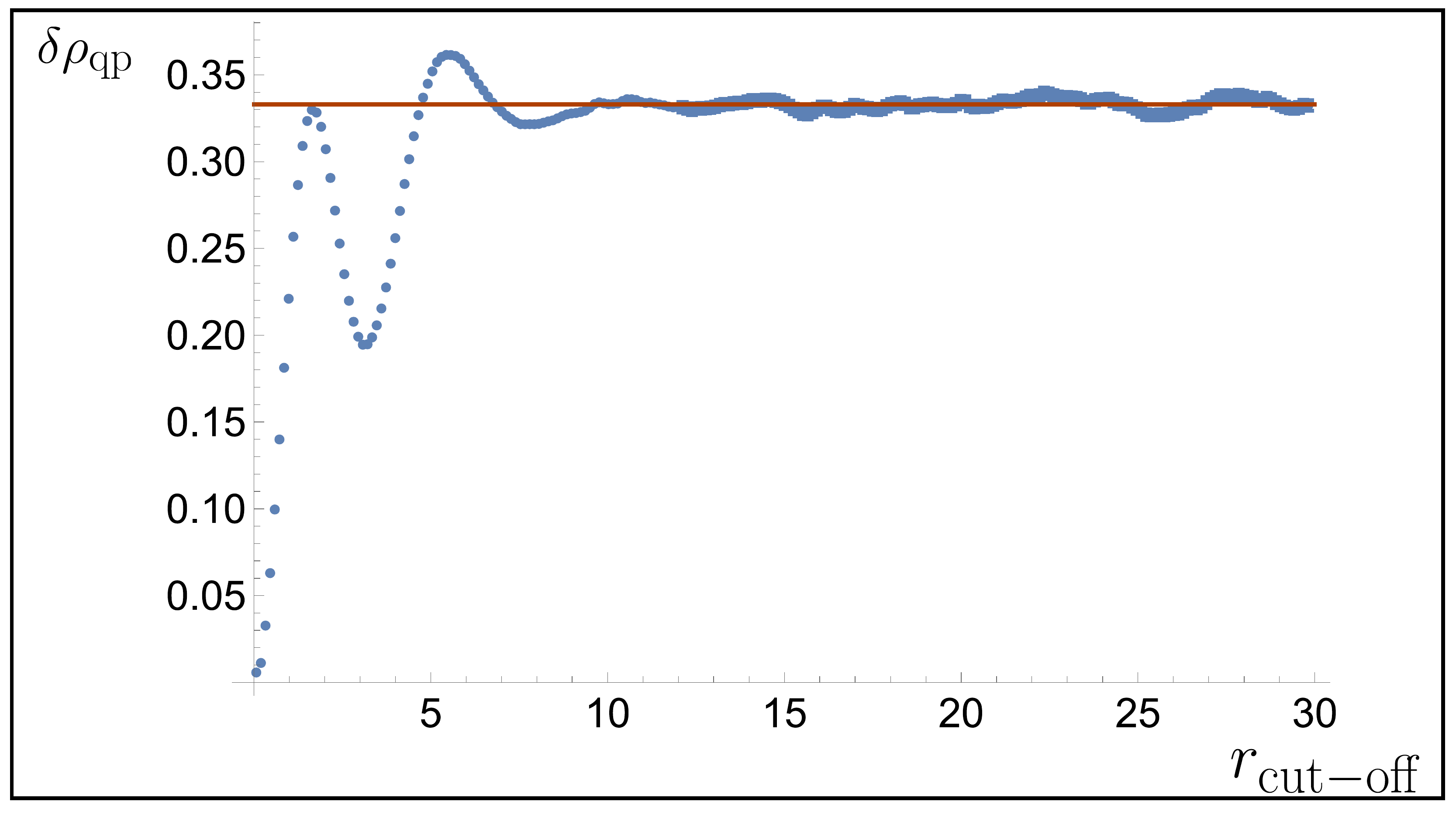}
\caption{ (Color online.) The charge $\delta\rho_{\mathrm{qp}}$ of the quasielectron (located at $\eta=0$) as a function of the cut-off radius $r_{\mathrm{cut-off}}$ (blue points), with the range of $r_{\mathrm{cut-off}}$ chosen so that boundary effects do not affect the result.  Saturation of $\delta\rho_{\mathrm{qp}}$ is clearly visible for $r_{\mathrm{cut-off}}\ge 12\ell$. The mean value for the charge in the range $r_{\mathrm{cut-off}}\in [12\ell, 30\ell]$ is then computed: $\delta\rho_{\mathrm{qp}} = 0.3330(30) e$  (solid line). Monte Carlo simulations averaged over more than $2\times 10^{10}$ equilibrated configurations.}
\label{fig:05}
\end{figure}

Using the method presented in \cite{KivelsonSchrieffer82} we can calculate the fractional charge of the quasielectron:
\begin{equation}
\label{eq:charge}
\delta \rho_{\mathrm{qp}} =2\pi \int_{0}^{r_{\mathrm{cut-off}}}
\left[\rho_{\mathrm{qp}}(r)-\rho_L(r)\right]r\,dr,
\end{equation}
with the cut-off radius $r_{\mathrm{cut-off}}$ chosen as discussed in the main text. The results for $N=400$ particles (obtained using Monte Carlo data, with $400$ grid points) are shown in Fig. \ref{fig:05}. Numerical integration is performed with the help of the trapezoidal method using the raw Monte Carlo data.

\section{Klein factors for the subspace bosonization}
\label{sec:OG}

We have defined $F_{M,N}^\dagger=\frac{1}{N+1}K_{M,N}$ and $\mathcal{F}_M^\dagger =\sum\limits_{N\ge 0}F^\dagger_{M,N}=\sum\limits_{N\ge 0}F_{M,N}^\dagger |\psi_M^N\rangle \langle\psi_M^N|$. Since $[\mathcal{O}_d, \overline{a}_r^\dagger]_{-}=\overline{a}^\dagger_{r+d}$ we immediately get
\begin{equation}
    [\mathcal{O}_d,F_{M,N}^\dagger]_{-}=\frac{(-1)^{MN}}{N+1}\sum\limits_{r\ge 0}\overline{a}_{r+d}^\dagger \widehat{S}^{\musSharp{}}_{MN-r}. 
\end{equation}
The operator $\overline{a}_{r+d}^\dagger \widehat{S}^{\musSharp{}}_{MN-r}$, for any $d>0$, annihilates given state or changes its angular momentum by $MN+d$, so as a result, $\langle \psi_M^{N+1}| [\mathcal{O}_d,F_{M,N}^\dagger]_{-}| \psi_M^N\rangle$=0, since the $N$-particle \mbox{$\frac{1}{M}$-Laughlin} state has an angular momentum equal to $J=\frac{MN(N-1)}{2}$. Next, we also have
\begin{equation}
    \begin{split}
        \langle\psi^{N+1}_M|[\mathcal{O}_d,\mathcal{F}_M^\dagger]_{-}|\psi_M^N\rangle& \\
& \hspace*{-3cm}  =\sum\limits_{N'\ge 0}\langle \psi_M^{N+1}|F^\dagger_{M,N}\left[\mathcal{O}_d, |\psi_{M}^{N'}\rangle\langle\psi_{M}^{N'}|\right]_{-}|\psi_{M}^N\rangle\\
& \hspace*{-3cm}  =\sum\limits_{N'\ge 0}\langle \psi_M^{N+1} |F_{M,N}^\dagger \mathcal{O}_d |\psi_M^{N'} \rangle\langle \psi_M^{N'}|\psi_{M}^N\rangle\\
& \hspace*{-3cm}  -\sum\limits_{N'\ge 0}\langle \psi_M^{N+1}|F^\dagger_{M,N'}|\psi_{M}^{N'}\rangle\langle\psi_{M}^{N'}|\mathcal{O}_d|\psi_{M}^N\rangle\\
& \hspace*{-3cm}  =\langle \psi_M^{N+1}|\mathcal{O}_d F^\dagger_{M,N}|\psi_M^N\rangle\|\psi_{M,N}\|^2 \\ 
& \hspace*{-1cm}-\sum\limits_{N'\ge 0}\langle\psi_M^{N+1}|\psi_M^{N'+1}\rangle\langle\psi_M^{N'}|\mathcal{O}_d|\psi_M^N\rangle\\
& \hspace*{-3cm}  =\langle\psi_M^{N+1}|\mathcal{O}_d |\psi_M^{N+1}\rangle\|\psi_{M}^N\|^2-\langle\psi_{M}^N|\mathcal{O}_d|\psi_M^N\rangle\|\psi_M^{N+1}\|^2 \\
& \hspace*{-3cm} =0,
    \end{split}
\end{equation}
for any $N\ge 0$ and any $d>0$.
Similar computations can be performed also for the commutators with $\mathcal{O}_d^\dagger$ as well as for the operator $\mathcal{F}_M=(\mathcal{F}_M^\dagger)^\dagger$.

\end{document}